\documentclass[9pt,twocolumn,twoside]{opticajnl}
\journal{opticajournal}

\setboolean{shortarticle}{true}
\usepackage{lineno}

\usepackage{subcaption}
\usepackage{amsmath}
\usepackage{amssymb}
\usepackage{graphicx}
\usepackage{booktabs}
\usepackage{amsfonts}

\setboolean{displaycopyright}{false}

\title{Optimal slit width for high-precision orbital angular momentum measurement using angular double-slit interferometry}

\author[1,2,3]{Yu Jian}
\author[1,2,3]{Xin Wang}
\author[1,2,3]{Jiyang Zhang}
\author[1,2,3]{Tao Chen}
\author[1,2,3]{Manpeng Chang}
\author[1,2,3]{Chen Liu}
\author[1,2,3,*]{Weimin Wang}

\affil[1]{Key Laboratory of Optoelectronic Technology and Systems, Ministry of Education, Chongqing University, Chongqing 400044, China}
\affil[2]{Defense Key Disciplines Laboratory of Novel Micro-Nano Devices and System Technology, Chongqing University, Chongqing 400044, China}
\affil[3]{College of Optoelectronic Engineering, Chongqing University, Chongqing 400044, China}

\affil[*]{wwm@cqu.edu.cn}

\begin{abstract}
We demonstrate an optimization of angular double-slit interferometry for accurate measurement of orbital angular momentum (OAM) of vortex beams. By scanning the dynamic double slits, the topological charge (TC) magnitude is directly determined from the oscillation frequency of the on-axis intensity. Based on repeated experimental investigations, we establish a critical criterion for slit width selection: to avoid phase truncation or period overlap, the angular width of each slit must exactly match the spiral phase period $2\pi/|l|$. Under this optimal condition, the interference pattern exhibits the highest visibility and the measurement error is minimized. Experiments for $l=5$, $10$, and $15$ are performed as representative examples, and the universality of this criterion is confirmed. Furthermore, by introducing an additional phase shift, the sign of the TC is unambiguously determined, as demonstrated for $l=10$ and $l=-15$. This simple, robust method provides a high-precision pathway for OAM metrology.
\end{abstract}

\begin{document}
\maketitle

\section{Introduction}

Vortex beams carrying orbital angular momentum are characterized by a helical phase front $\exp(i\ell\varphi)$ \cite{Allen1992}, where $\varphi$ is the azimuthal angle and $\ell$ is the topological charge. Such beams possess a phase singularity on the propagation axis, resulting in a doughnut-shaped intensity profile and carrying an OAM of $\ell\hbar$ per photon \cite{Allen1992, Yao2011}. OAM beams have found widespread applications in optical manipulation, quantum information processing, high-capacity optical communications, and super-resolution imaging \cite{Yao2011, Shen2019}. The value of $\ell$ can be any integer (or even fractional), offering an unbounded state space, making accurate TC determination a fundamental task \cite{Zhu2016}.

Numerous methods have been developed to measure the TC \cite{Yao2011, Shen2019}. Traditional interferometric approaches \cite{Harris1994, Padgett1996} require a stable reference beam. Diffraction-based techniques using various apertures \cite{Hickmann2010, Guo2009, Berkhout2008, Berkhout2009, Zhao2020} are simpler but typically limited to small TCs. Double-slit interference in Cartesian coordinates \cite{Sztul2006, Zhou2014b} suffers from degraded resolution for high-order modes. Mode converters \cite{Beijersbergen1993, Zhou2016} and log-polar transformation \cite{Berkhout2010, Berkhout2011, Mirhosseini2013, Lavery2012} enable efficient sorting but involve complex optics.

Dynamic angular double-slit interferometry has emerged as a simple and versatile technique \cite{Liu2013, Fu2015, Zhu2016}. Two sector-shaped slits of equal angular width $\alpha$ and variable relative angle $\varphi$ are illuminated by the vortex beam. By fixing the observation point at the angular bisector, the far-field intensity follows $I \propto 1+\cos(\ell\varphi)$ when the optical path difference is zero. Scanning $\varphi$ yields an oscillatory curve whose frequency directly gives $|\ell|$, and an additional phase shift $\theta$ on one slit allows sign determination from the rotation direction of the $I(\varphi)$ curve \cite{Fu2015}. This method has been extended to fractional TCs \cite{Zhu2016, Deng2019} and perfect vortex beams \cite{Zhao2021}.

Despite its simplicity, the measurement accuracy critically depends on the angular width $\alpha$. Previous implementations typically used a fixed narrow slit ($8^\circ$ \cite{Fu2015, Zhu2016} or $1$ mm in the spatial domain \cite{Zhou2014a}) to avoid phase averaging, but the influence of $\alpha$ on the fidelity of $I(\varphi)$ has not been systematically investigated. An excessively narrow slit reduces transmitted intensity and truncates the spiral phase, leading to low visibility. An overly wide slit covers more than one $2\pi$ phase period, causing multiple lobes to overlap and resulting in missing peaks and erroneous TC identification \cite{Zhou2014a}. Therefore, establishing an optimal slit width is essential for high-precision OAM metrology.

In this work, we systematically study the role of slit width in dynamic ADS interferometry. We experimentally verify that the optimal slit width for an integer vortex beam is exactly $\alpha_{\mathrm{opt}} = 2\pi/|\ell|$, i.e., each slit spans precisely one full $2\pi$ phase cycle. Under this condition, the interference signal $I(\varphi)$ most closely matches the ideal cosine function, maximizing visibility and minimizing error. We comprehensively validate this condition for $\ell=5$, $10$, and $15$, and demonstrate sign determination for both positive and negative TCs ($\ell=10$ and $\ell=-15$). Our findings provide a clear guideline for optimizing dynamic angular double-slit interferometry.

\section{Principle and experimental setup}

\begin{figure}[ht]
\centering
\includegraphics[width=0.8\linewidth]{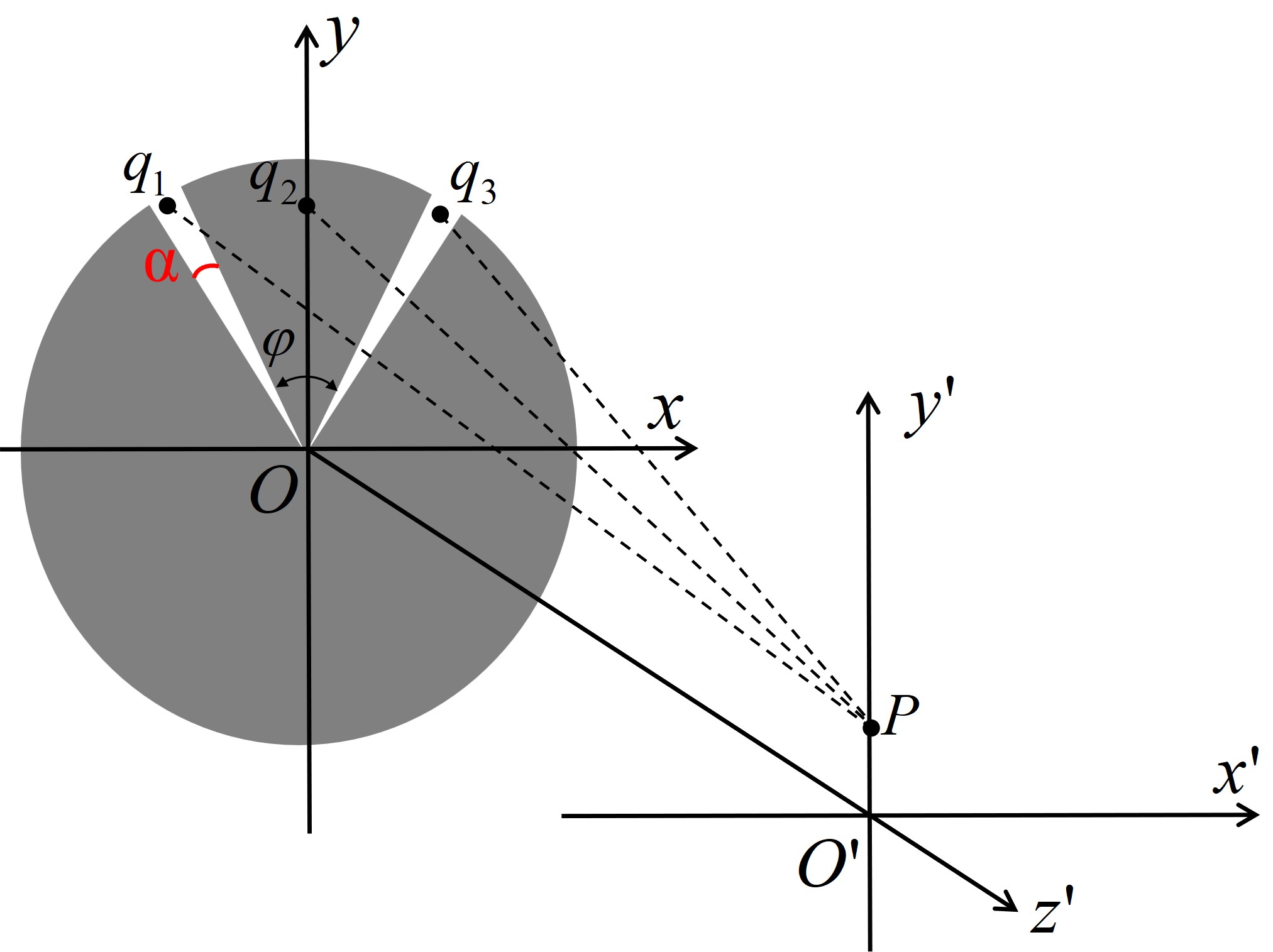}
\caption{Schematic of dynamic angular double-slit interferometry. Slit angular width $\alpha$, relative angle $\varphi$. Points $q_1$, $q_2$, $q_3$ on the mask; $P$ is the far-field observation point satisfying $q_1P = q_2P$. $\theta$ denotes an extra phase shift applied to one slit.}
\label{fig:ads_schema}
\end{figure}

The dynamic angular double slit consists of two coplanar sector-shaped slits of equal angular width $\alpha$, with a variable relative angle $\varphi$ (Fig.~\ref{fig:ads_schema}). When an integer vortex beam (phase factor $\exp(i\ell\phi)$) is normally incident, the light transmitted through the slits interferes in the far field. Under the condition that the observation point $P$ satisfies $q_1P = q_2P$, the intensity can be approximated as \cite{Fu2015}:
\begin{equation}
I \propto 1 + \cos(\ell \varphi + \theta),
\label{eq:interference}
\end{equation}
where $\varphi_1$ and $\varphi_2$ are the azimuthal centers of the two slits, $\varphi = \varphi_2 - \varphi_1$, and $\theta$ is an additional phase applied to one slit. Scanning $\varphi$ yields an oscillatory $I(\varphi)$ curve with frequency $|\ell|$, allowing direct determination of the TC magnitude. The sign of $\ell$ is obtained by comparing curves with $\theta=0$ and $\theta=\pi/2$: for $\ell>0$ the curve rotates clockwise, for $\ell<0$ counterclockwise. The angular shift of the $I(\varphi)$ curve, denoted as $\delta$, is given by
\begin{equation}
\delta = \frac{\theta}{\ell},
\label{eq:rotation}
\end{equation}
where $\theta$ is the additional phase shift applied to one slit. The direction of this shift (clockwise for $\ell>0$, counterclockwise for $\ell<0$) unambiguously reveals the sign of the topological charge.

\begin{figure}[ht]
\centering
\includegraphics[width=0.8\linewidth]{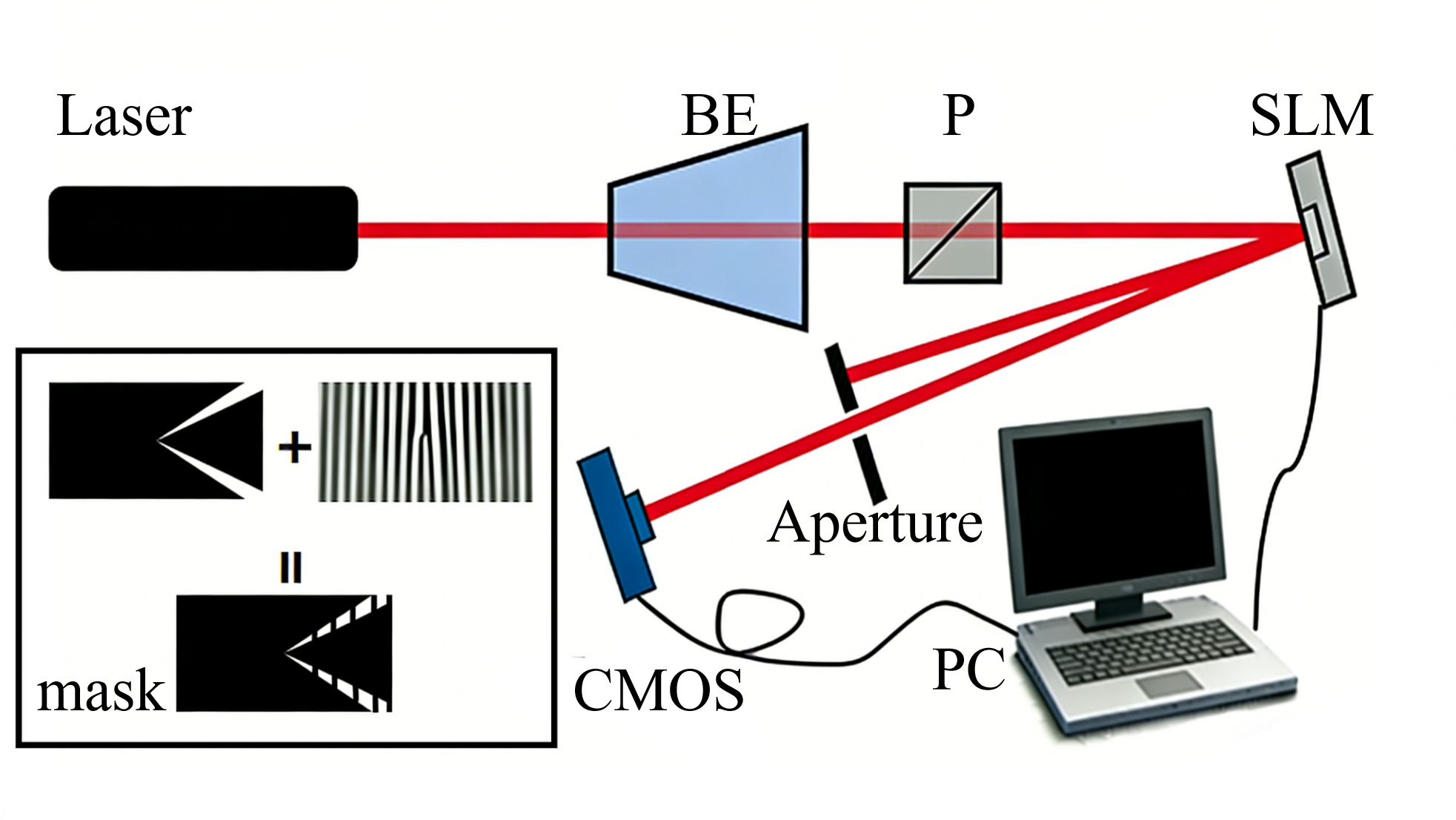}
\caption{Schematic of the experimental setup. BE, beam expander; P, polarizer; SLM, spatial light modulator. Inset: typical phase mask loaded on the SLM.}
\label{fig:exp_setup}
\end{figure}

The experimental setup is shown in Fig.~\ref{fig:exp_setup}. A He-Ne laser (632.8 nm) beam is expanded, collimated, and converted to linear polarization by a polarizer. The beam is incident onto a reflective liquid-crystal spatial light modulator (SLM, FSLM-2K70-P02, 1920$\times$1080 pixels, 8.0 $\mu$m pitch). The SLM is loaded with computer-generated phase masks that encode both a dynamic angular double slit and a vortex beam. The first-order diffracted beam is selected by an aperture, and the interference pattern is recorded by a CMOS camera. By integrating the intensity over the central region of the pattern, the intensity curve $I(\varphi)$ is obtained as a function of the angular separation $\varphi$ between the two slits.

To ensure statistical reliability, all measurements were repeated multiple times under identical conditions. For each fixed $\ell$ and $\alpha$, the intensity curve $I(\varphi)$ was recorded over the full $0^\circ$--$360^\circ$ range of $\varphi$ with a step size of $1^\circ$. The multiple measurements were then averaged before further fitting.

\section{Simulation and Experimental Results}

\subsection{Interference patterns at different angular separations}

\begin{figure}[htbp]
\centering
\begin{subfigure}[b]{0.22\linewidth}
\includegraphics[width=\linewidth]{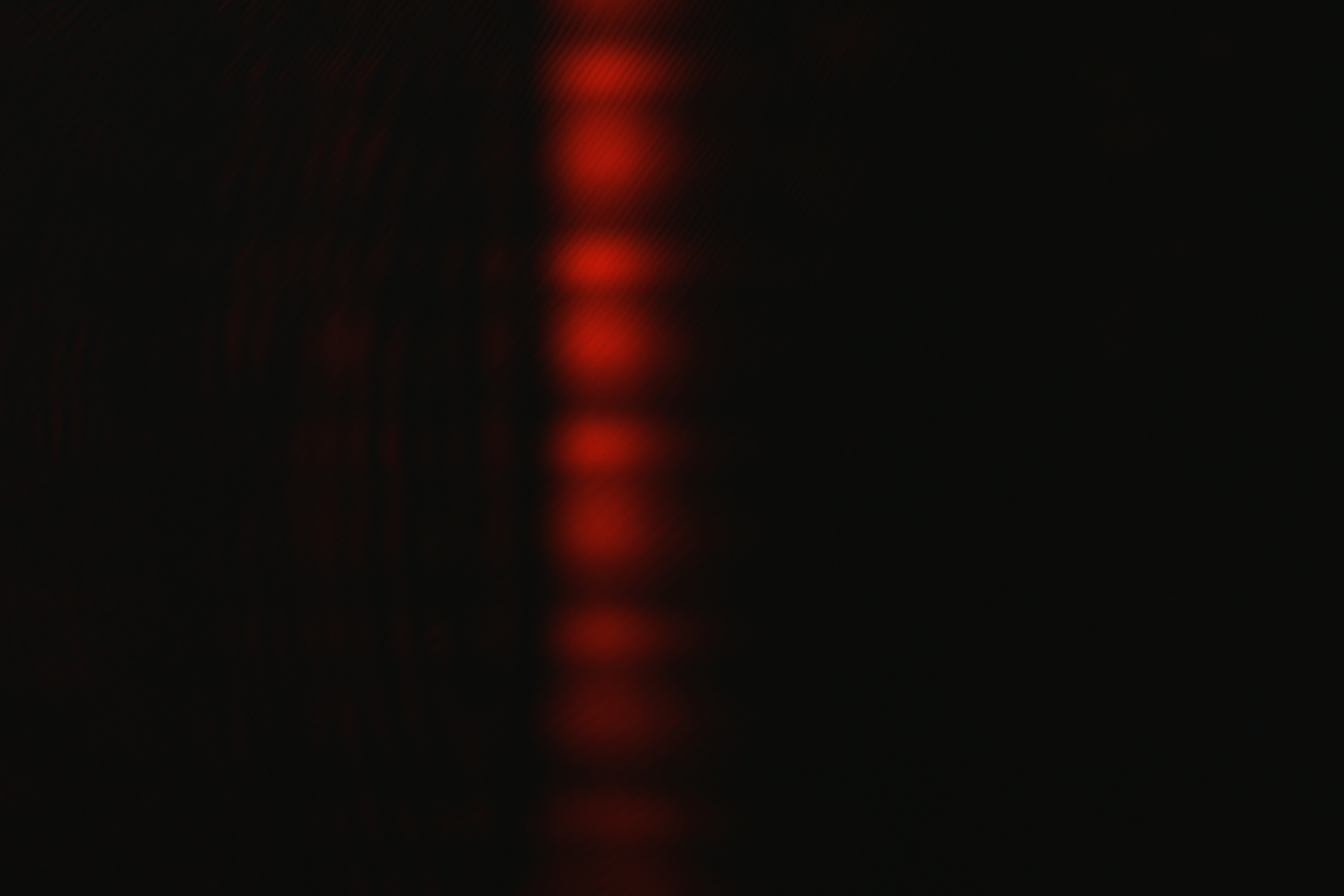}
\caption{$\varphi=0^\circ$}
\end{subfigure}
\hfill
\begin{subfigure}[b]{0.22\linewidth}
\includegraphics[width=\linewidth]{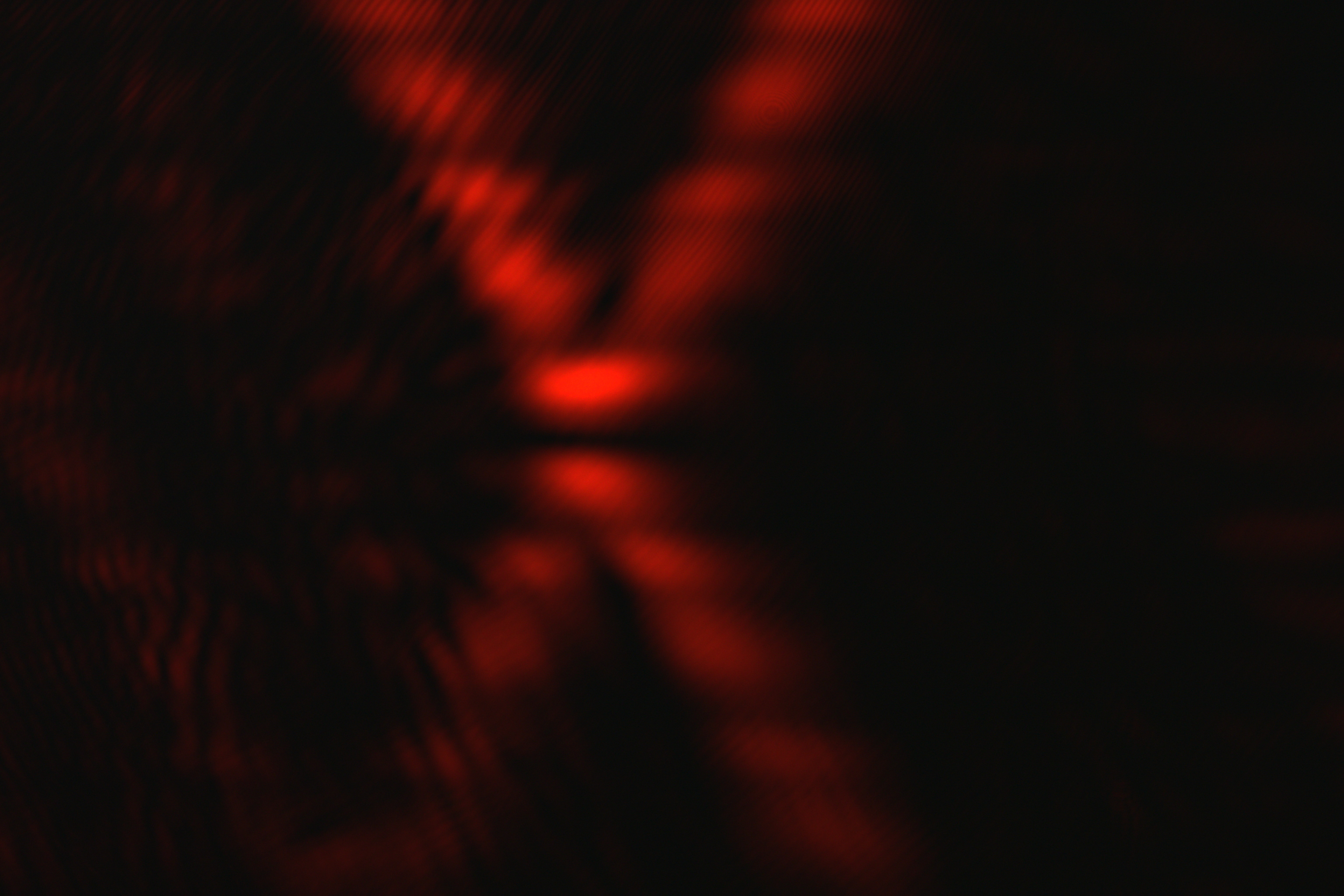}
\caption{$\varphi=60^\circ$}
\end{subfigure}
\hfill
\begin{subfigure}[b]{0.22\linewidth}
\includegraphics[width=\linewidth]{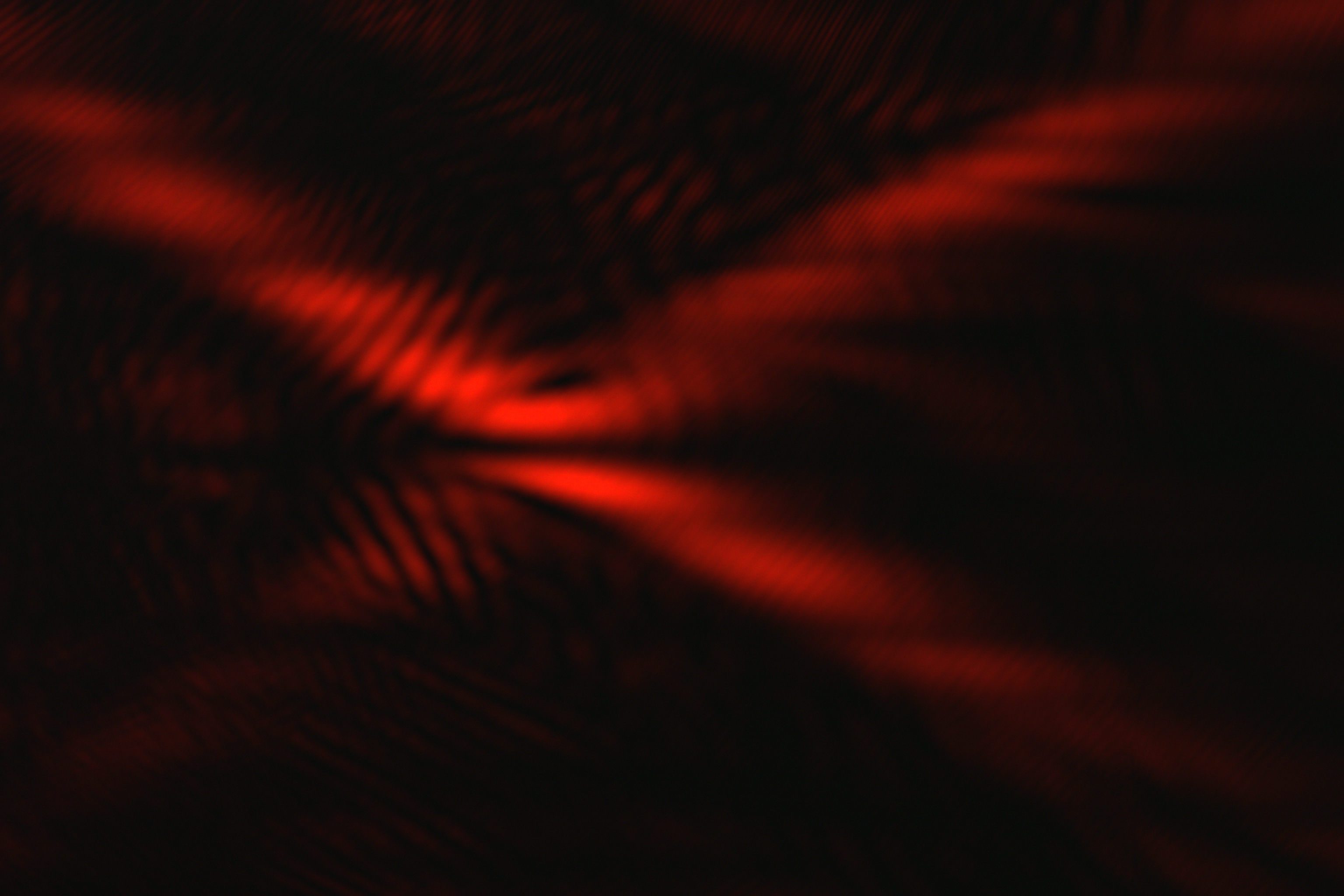}
\caption{$\varphi=120^\circ$}
\end{subfigure}
\hfill
\begin{subfigure}[b]{0.22\linewidth}
\includegraphics[width=\linewidth]{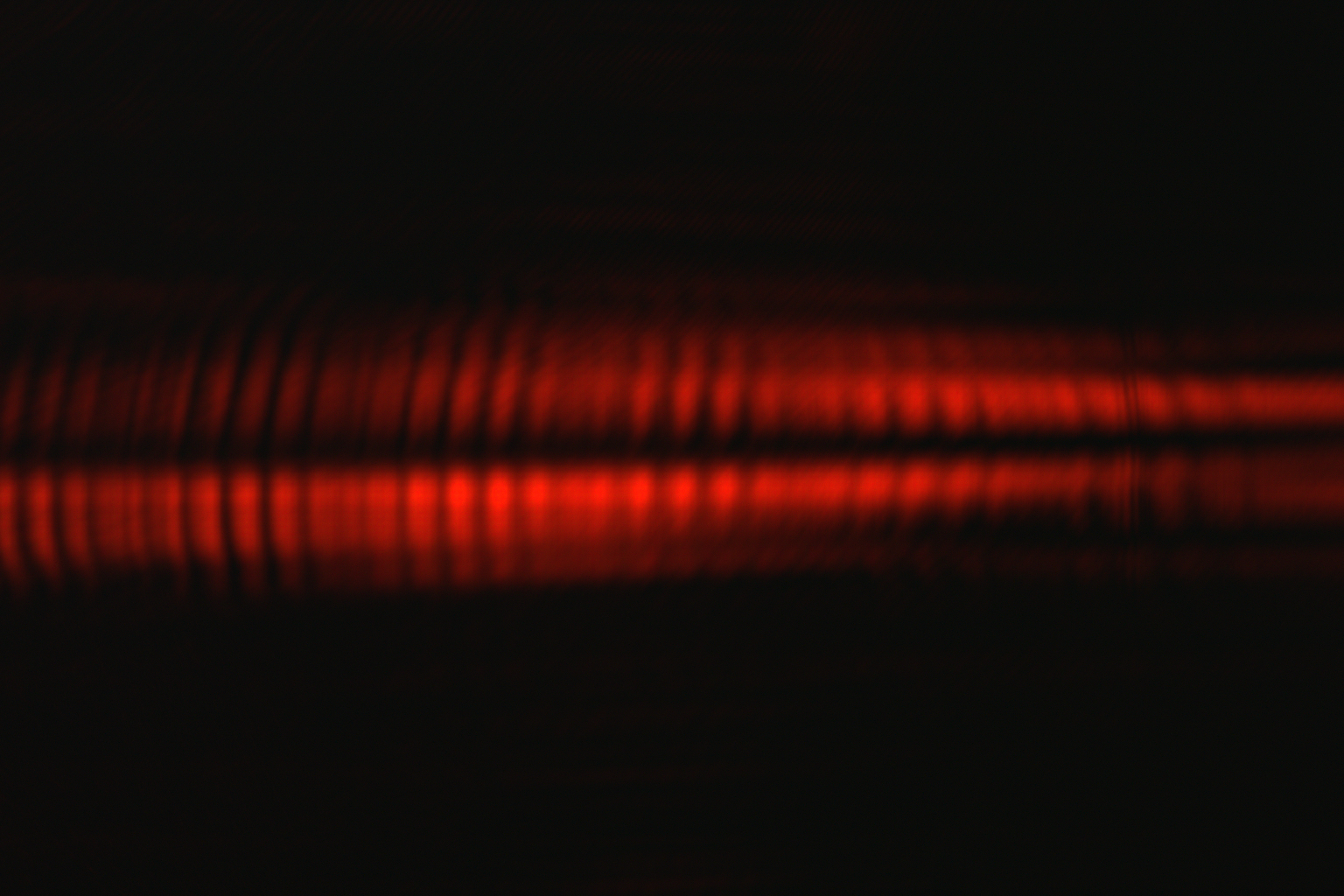}
\caption{$\varphi=180^\circ$}
\end{subfigure}

\vspace{0.5em}
\begin{subfigure}[b]{0.22\linewidth}
\includegraphics[width=\linewidth]{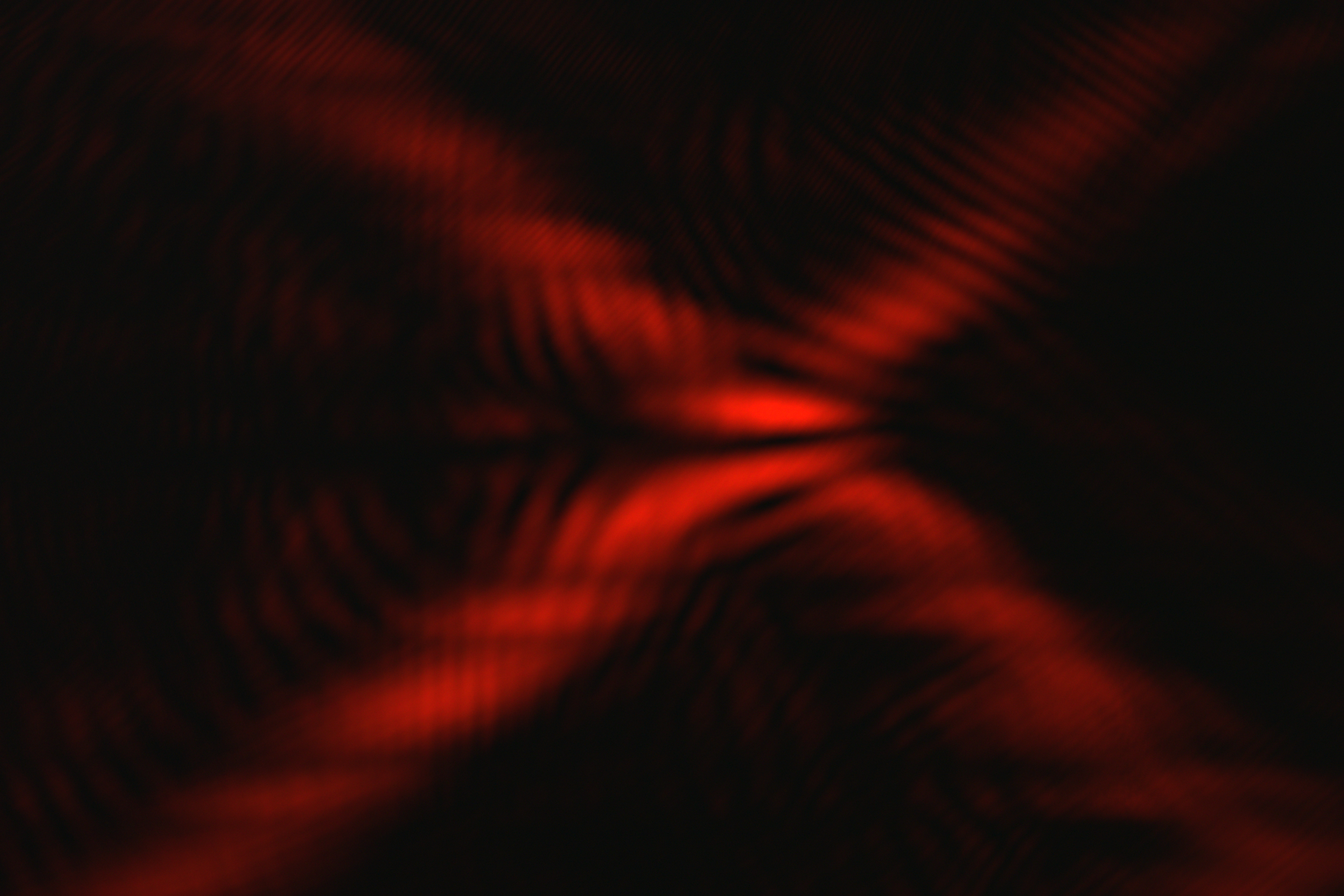}
\caption{$\varphi=240^\circ$}
\end{subfigure}
\hfill
\begin{subfigure}[b]{0.22\linewidth}
\includegraphics[width=\linewidth]{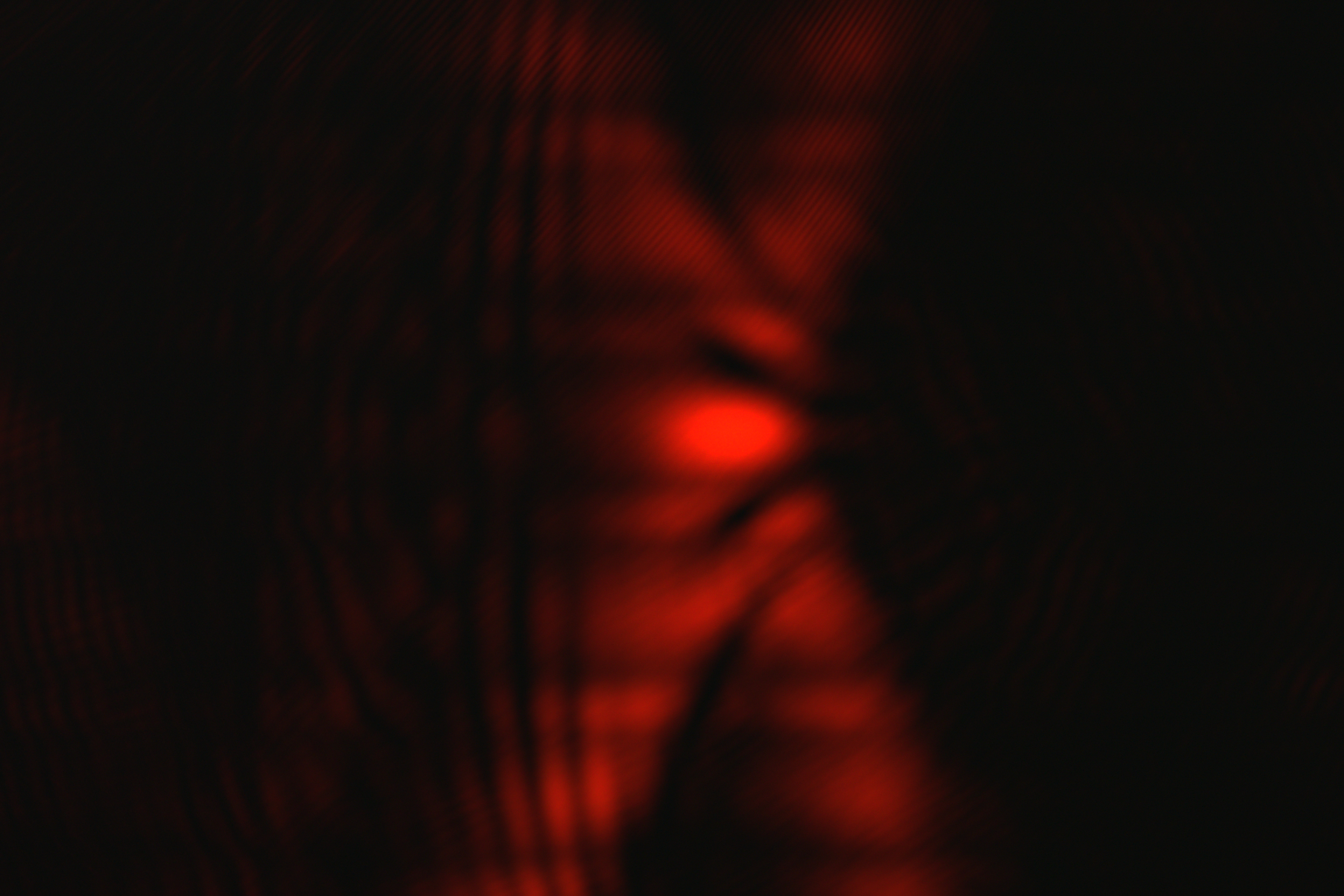}
\caption{$\varphi=300^\circ$}
\end{subfigure}
\hfill
\begin{subfigure}[b]{0.22\linewidth}
\includegraphics[width=\linewidth]{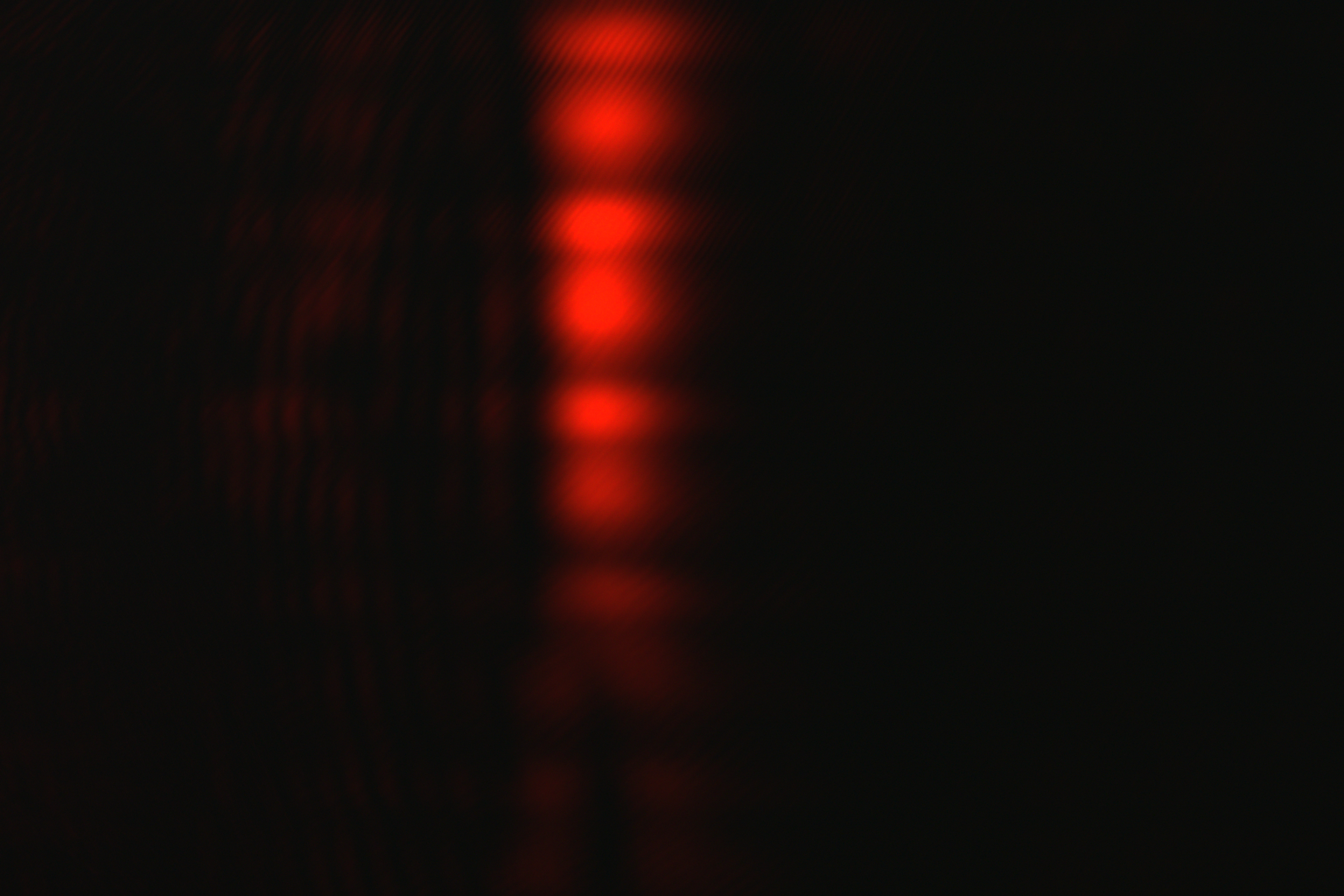}
\caption{$\varphi=360^\circ$}
\end{subfigure}
\caption{Far-field interference patterns of the angular double slit at various relative angles $\varphi$ for a vortex beam with topological charge $l = 10$ and optimal slit width $\alpha = 36^\circ$. As $\varphi$ increases, the central intensity undergoes periodic oscillations due to the phase difference imprinted by the helical phase front.}
\label{fig:interference_patterns}
\end{figure}

For a vortex beam with topological charge $l = 10$ modulated by the SLM, the first-order diffracted beam is selected by an adjustable aperture. Figure~\ref{fig:interference_patterns} shows the far-field interference patterns of the angular double slit as a function of the relative angle $\varphi$ between the two slits. When $\varphi$ is continuously varied, the intensity at the center of the pattern exhibits periodic bright–dark alternations. This behavior originates from the phase difference introduced by the helical phase term of the vortex beam between the two slits. The oscillation period is directly determined by the magnitude of the topological charge $l$ of the incident beam. By recording the central intensity and analyzing its dependence on $\varphi$, the absolute value of $l$ can be accurately determined.

\begin{figure}[htbp!]
\centering
\begin{subfigure}[b]{0.48\linewidth}
    \includegraphics[width=\linewidth]{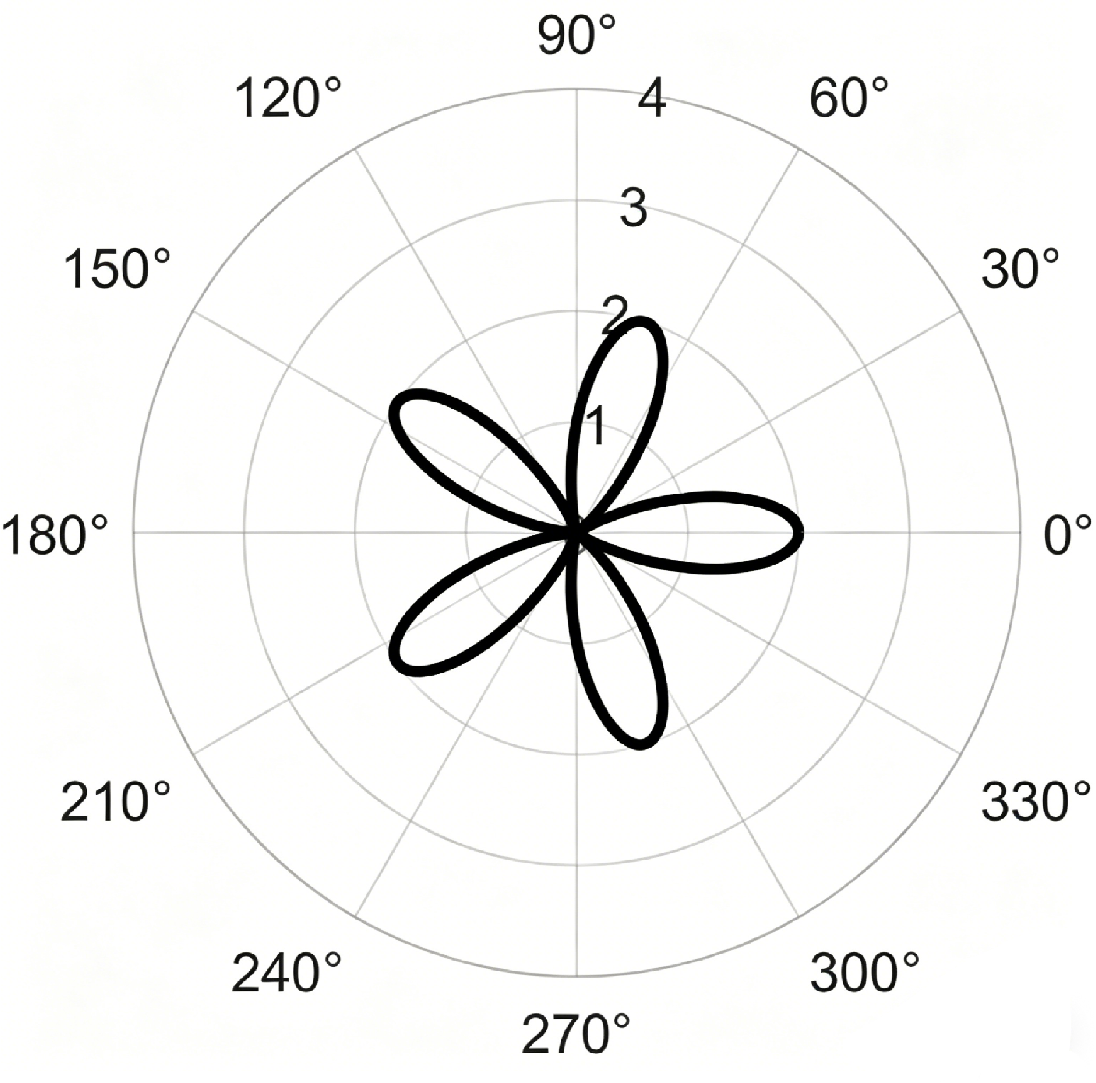}
    \caption{Theory, $\alpha=72^\circ$}
\end{subfigure}
\hfill
\begin{subfigure}[b]{0.48\linewidth}
    \includegraphics[width=\linewidth]{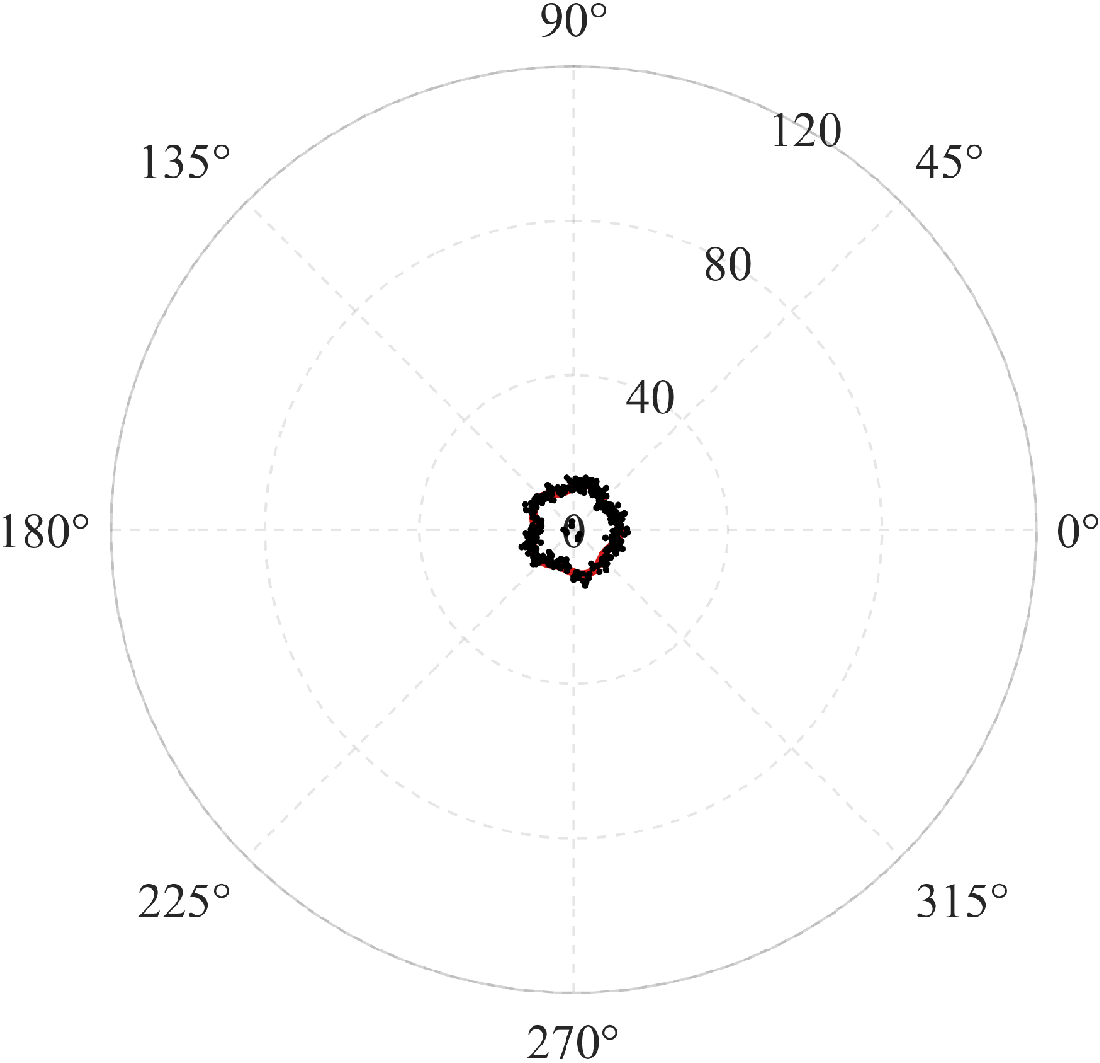}
    \caption{Exp., $\alpha=8^\circ$}
\end{subfigure}

\vspace{0.5em}
\begin{subfigure}[b]{0.48\linewidth}
    \includegraphics[width=\linewidth]{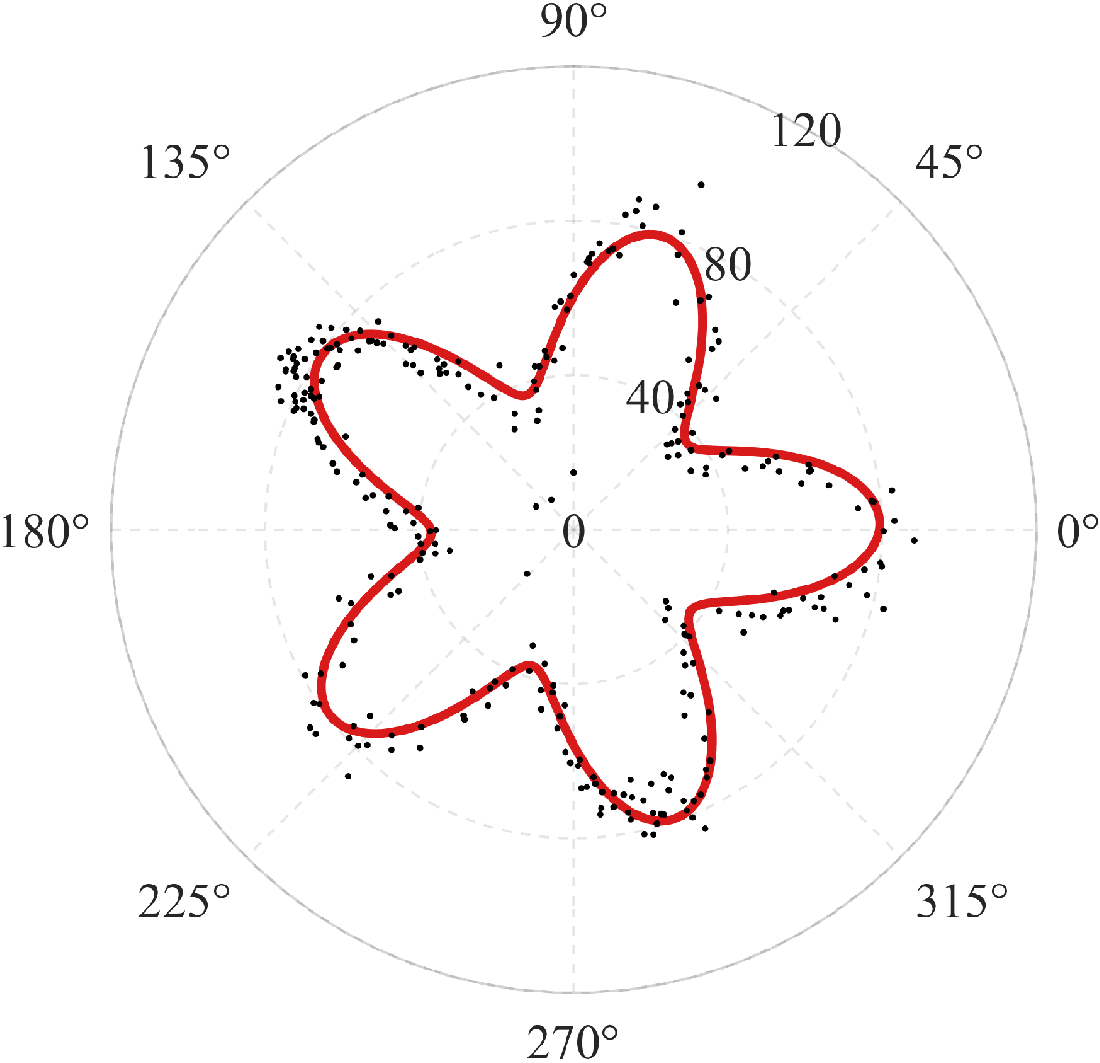}
    \caption{Exp., $\alpha=50^\circ$}
\end{subfigure}
\hfill
\begin{subfigure}[b]{0.48\linewidth}
    \includegraphics[width=\linewidth]{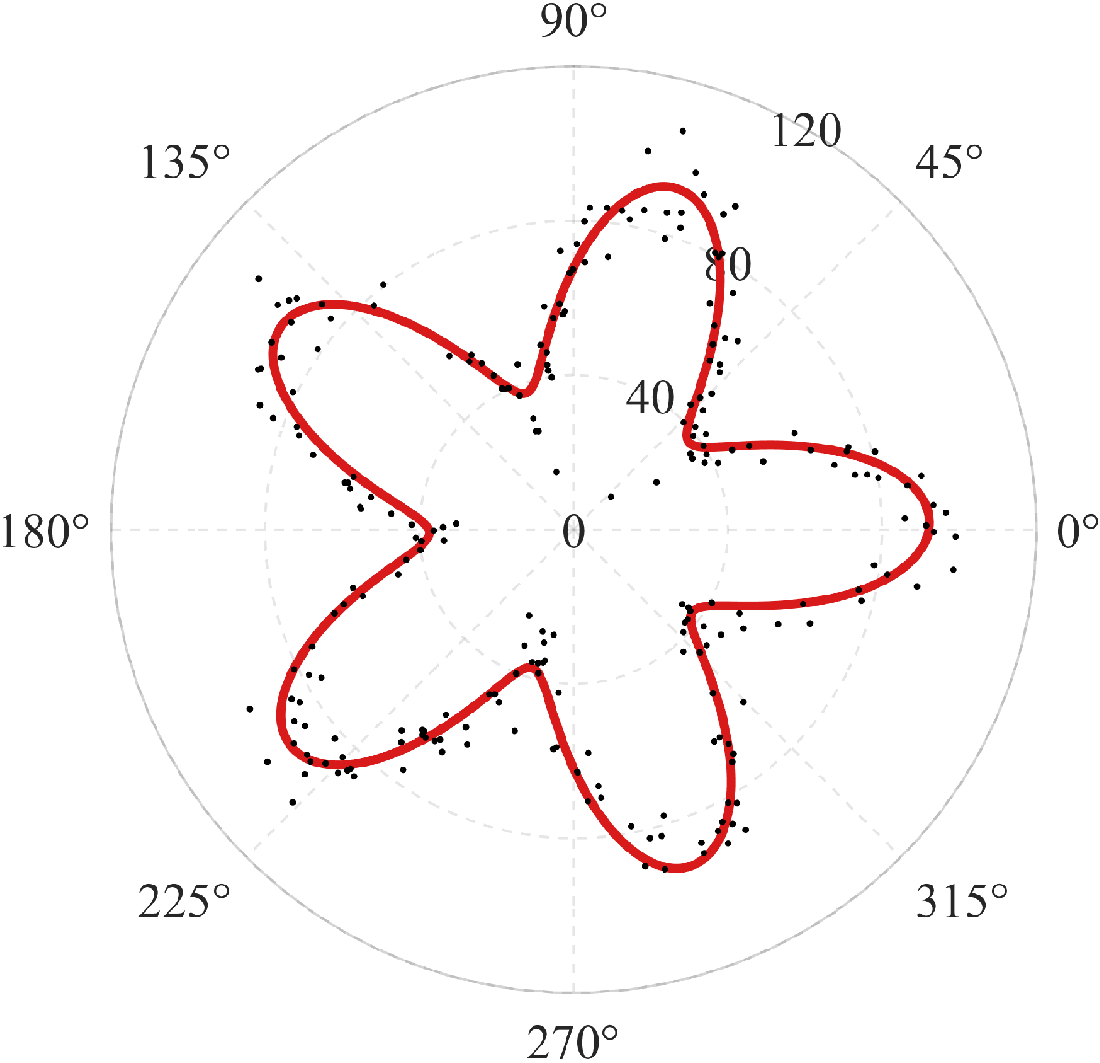}
    \caption{Exp., $\alpha=72^\circ$}
\end{subfigure}

\vspace{0.5em}
\begin{subfigure}[b]{0.48\linewidth}
    \includegraphics[width=\linewidth]{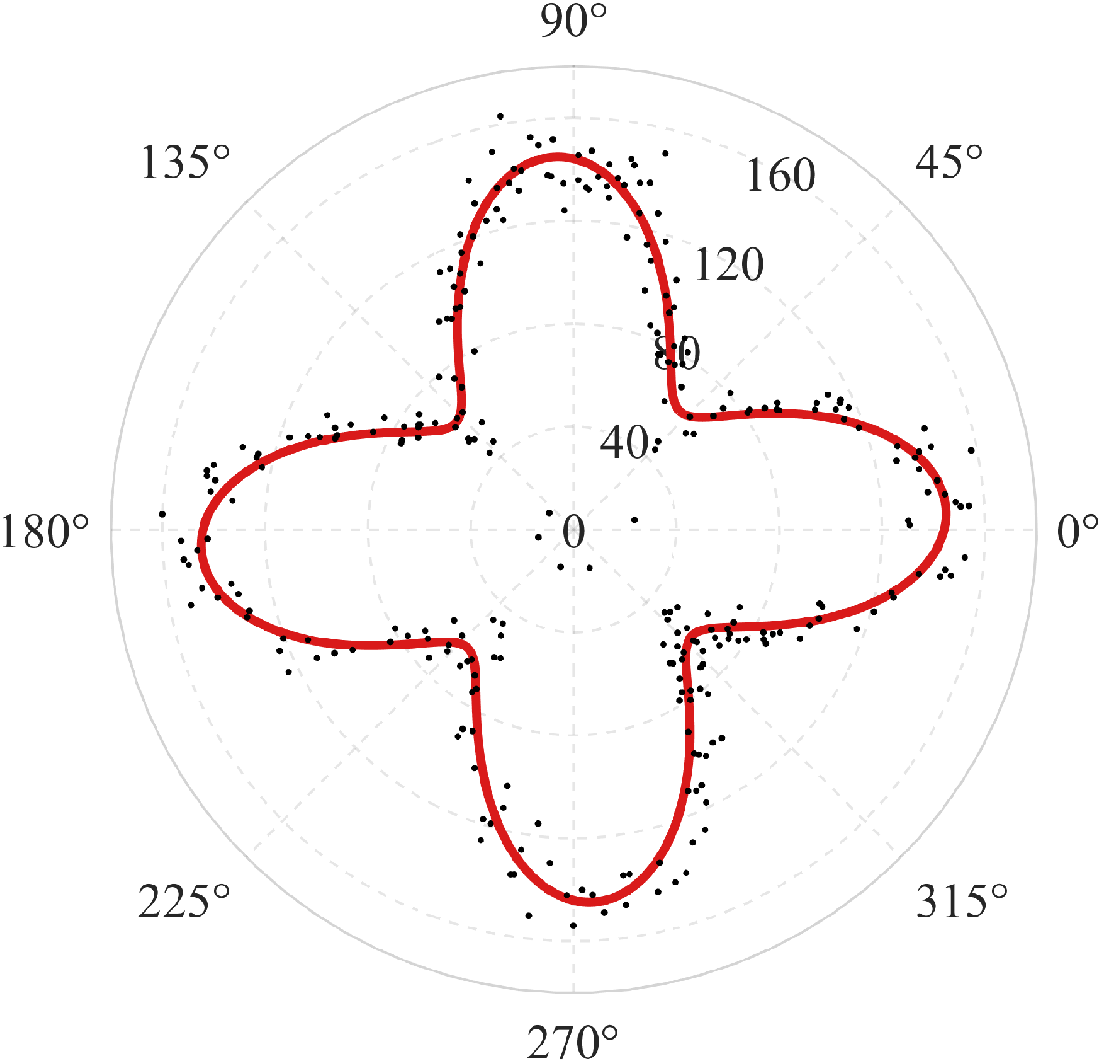}
    \caption{Exp., $\alpha=80^\circ$}
\end{subfigure}
\hfill
\begin{subfigure}[b]{0.48\linewidth}
    \includegraphics[width=\linewidth]{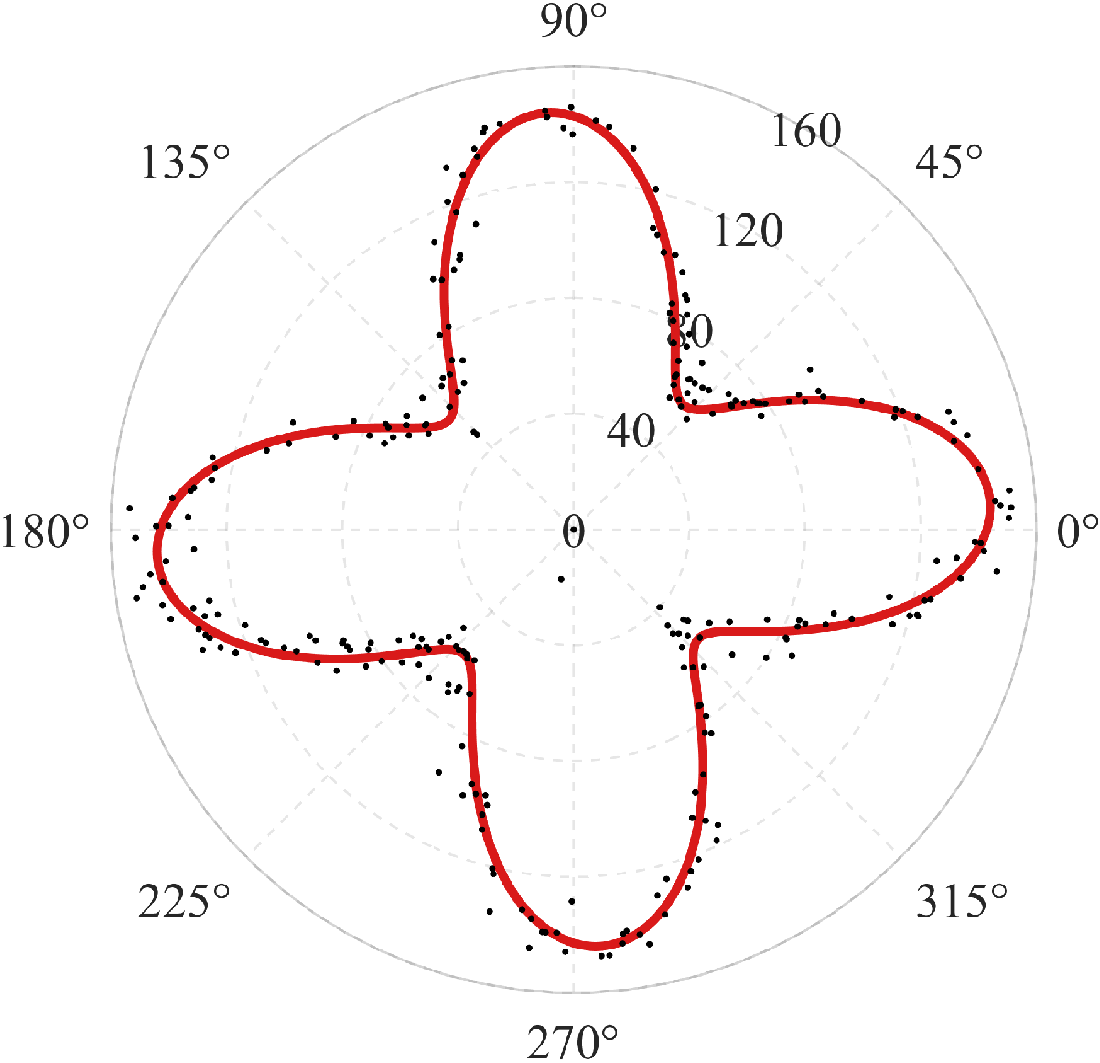}
    \caption{Exp., $\alpha=90^\circ$}
\end{subfigure}
\caption{Polar plots $I(\varphi)$ for $l=5$: (a) theoretical ideal curve at optimal slit width $72^\circ$; (b)--(f) experimental results for different slit widths. The optimal width yields five clear lobes.}
\label{fig:polar_l5}
\end{figure}

\begin{figure}[htbp!]
\centering
\begin{subfigure}{0.48\linewidth}
\includegraphics[width=\linewidth]{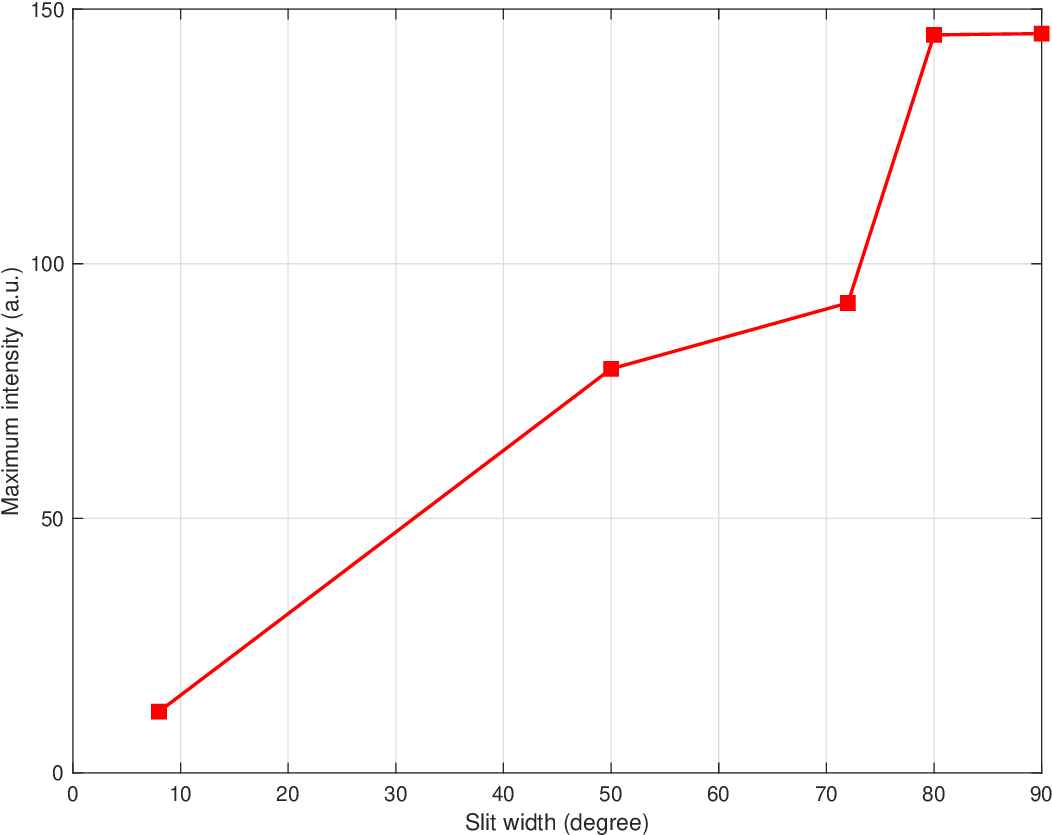}
\caption{Maximum intensity vs. $\alpha$}
\end{subfigure}
\hfill
\begin{subfigure}{0.48\linewidth}
\includegraphics[width=\linewidth]{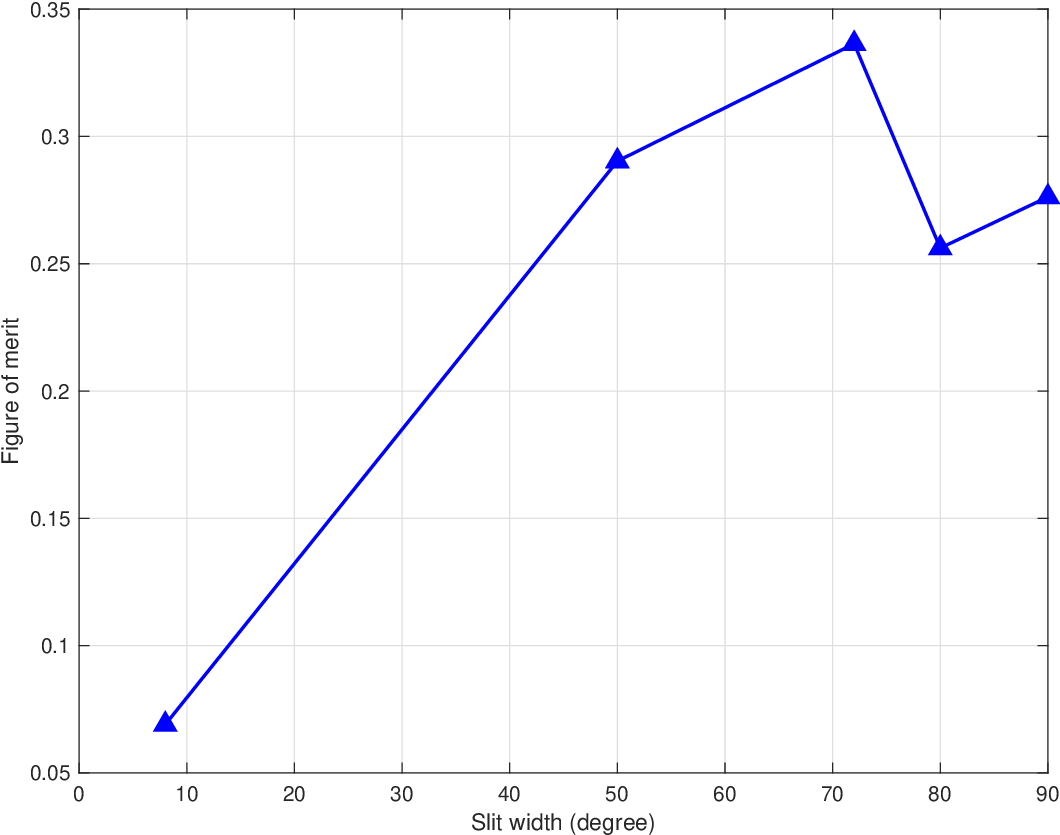}
\caption{FOM vs. $\alpha$}
\end{subfigure}
\caption{Quantitative analysis for $l=5$. (a) Maximum intensity as a function of slit width; (b) figure of merit as a function of slit width. The FOM peaks at the optimal width $\alpha=72^\circ$.}
\label{fig:quant_l5}
\end{figure}

\begin{figure}[htbp!]
\centering
\begin{subfigure}[b]{0.48\linewidth}
    \includegraphics[width=\linewidth]{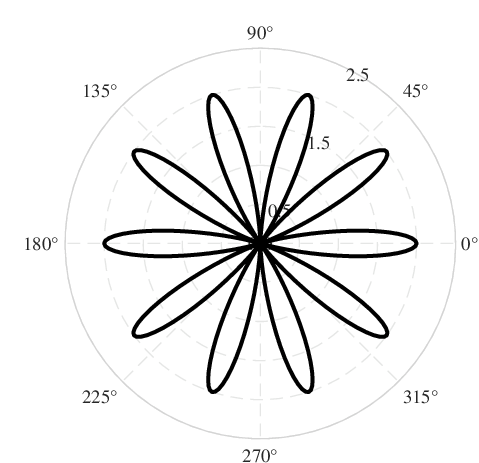}
    \caption{Theory, $\alpha=36^\circ$}
\end{subfigure}
\hfill
\begin{subfigure}[b]{0.48\linewidth}
    \includegraphics[width=\linewidth]{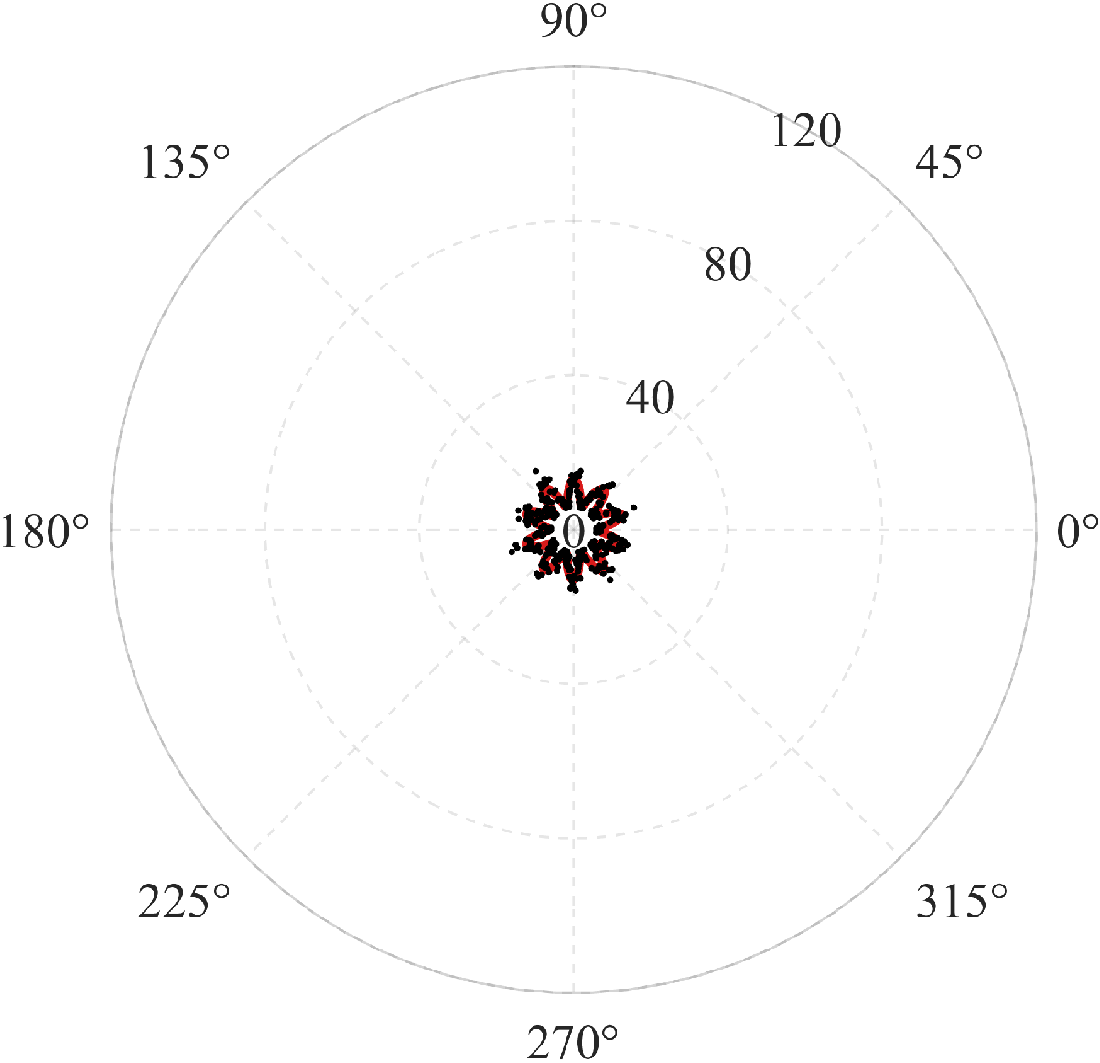}
    \caption{Exp., $\alpha=8^\circ$}
\end{subfigure}

\vspace{0.5em}
\begin{subfigure}[b]{0.48\linewidth}
    \includegraphics[width=\linewidth]{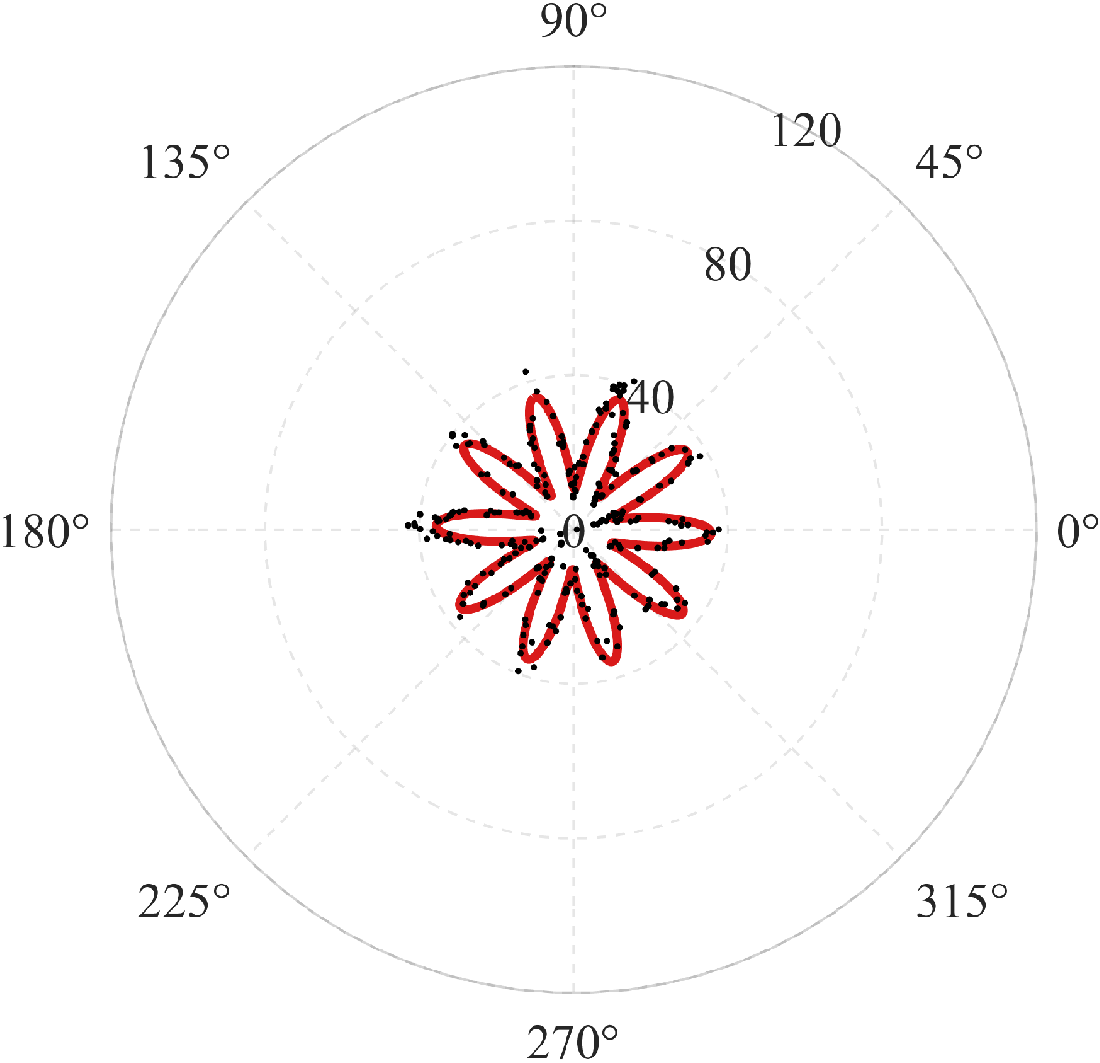}
    \caption{Exp., $\alpha=15^\circ$}
\end{subfigure}
\hfill
\begin{subfigure}[b]{0.48\linewidth}
    \includegraphics[width=\linewidth]{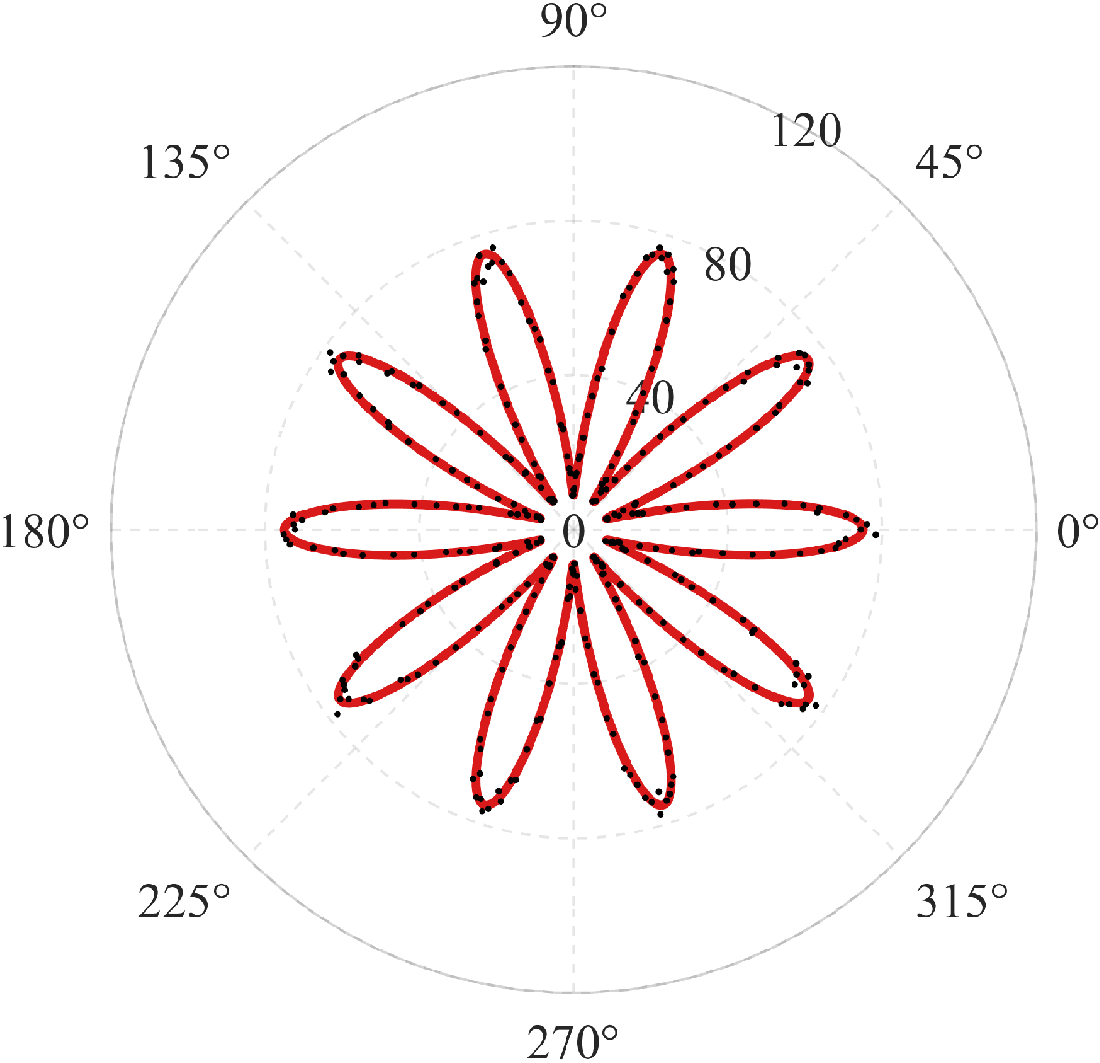}
    \caption{Exp., $\alpha=36^\circ$}
\end{subfigure}

\vspace{0.5em}
\begin{subfigure}[b]{0.48\linewidth}
    \includegraphics[width=\linewidth]{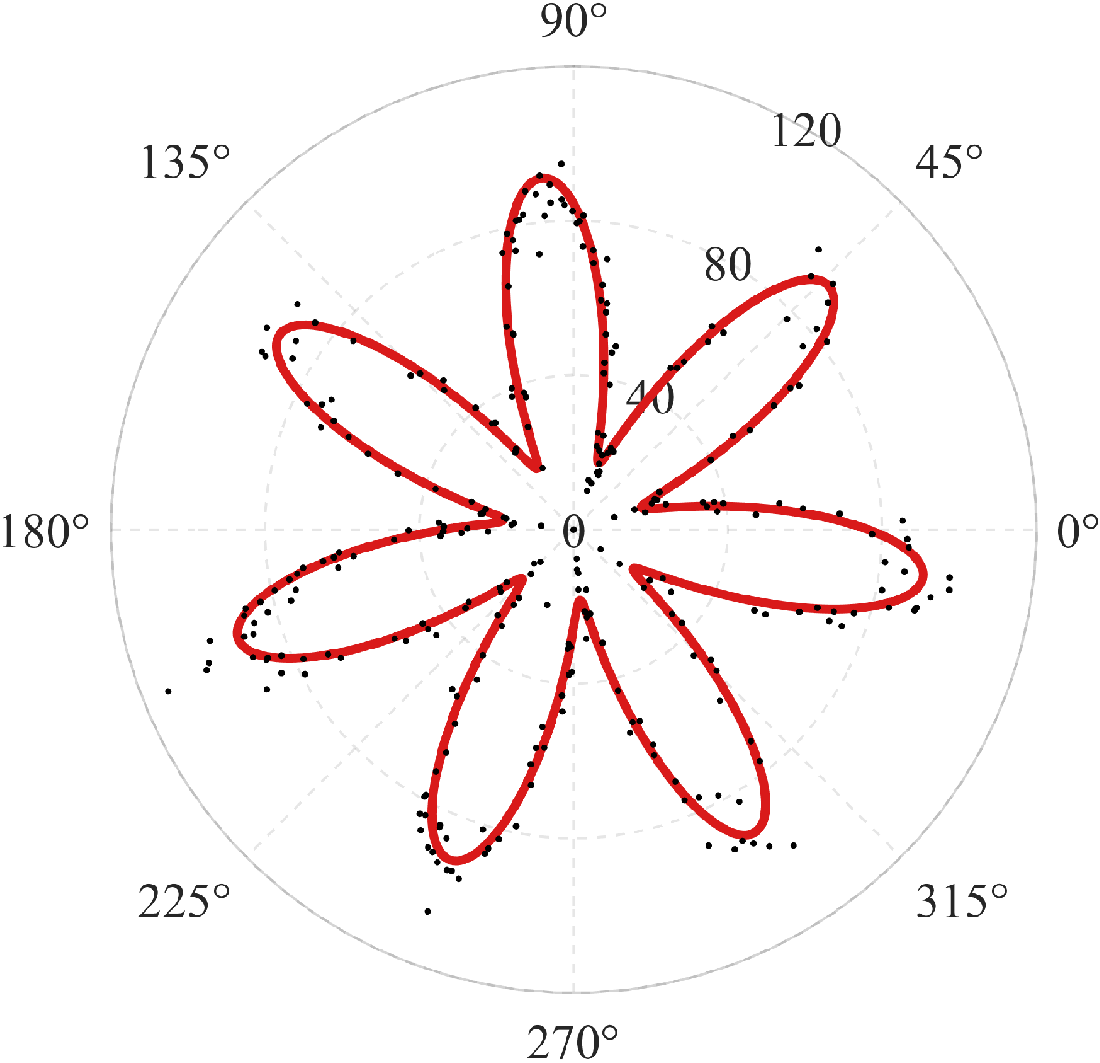}
    \caption{Exp., $\alpha=50^\circ$}
\end{subfigure}
\hfill
\begin{subfigure}[b]{0.48\linewidth}
    \includegraphics[width=\linewidth]{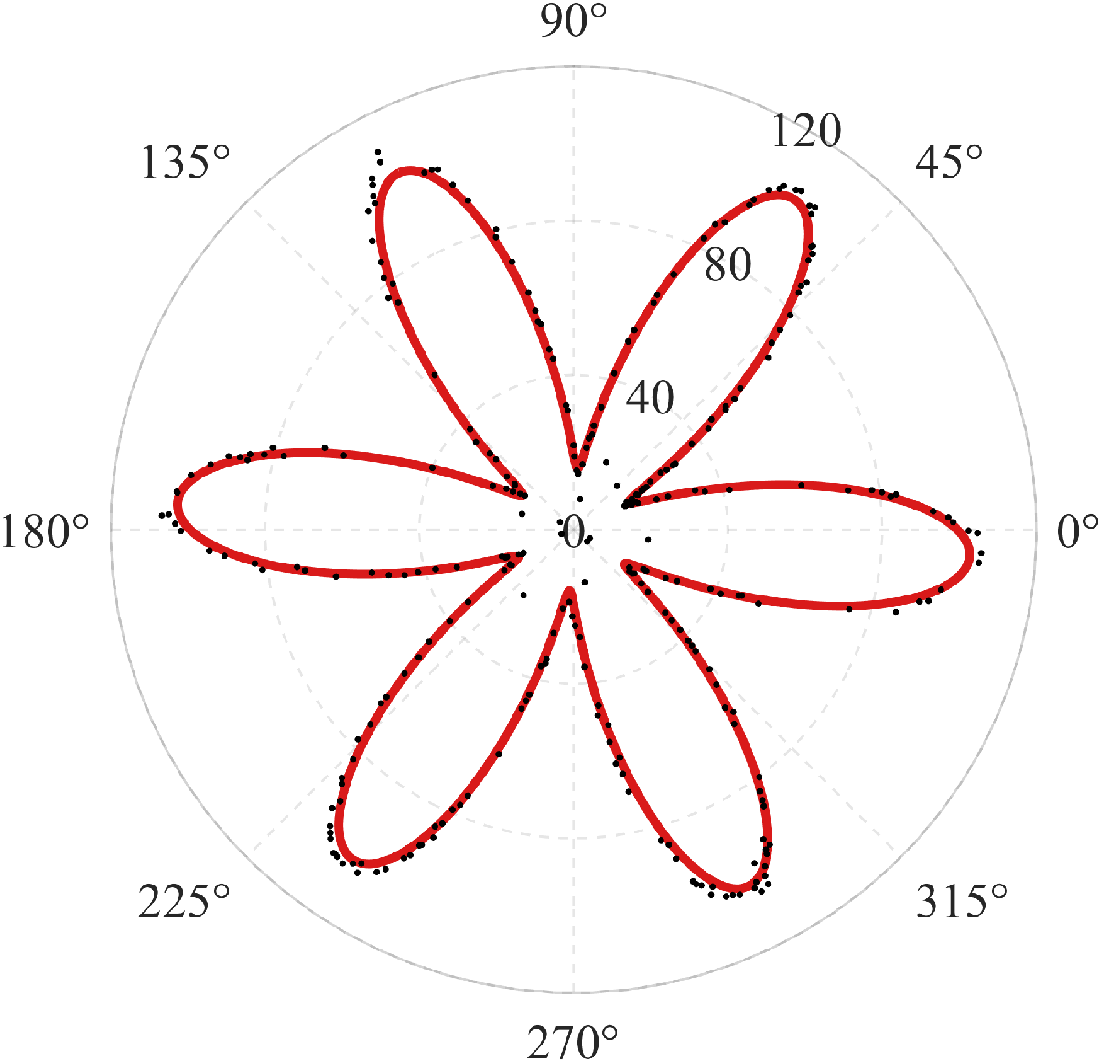}
    \caption{Exp., $\alpha=60^\circ$}
\end{subfigure}
\caption{Polar plots $I(\varphi)$ for $l=10$: (a) theoretical ideal curve at optimal slit width $36^\circ$; (b)--(f) experimental results. Optimal width gives ten clear lobes.}
\label{fig:polar_l10}
\end{figure}

\begin{figure}[htbp!]
\centering
\begin{subfigure}{0.48\linewidth}
\includegraphics[width=\linewidth]{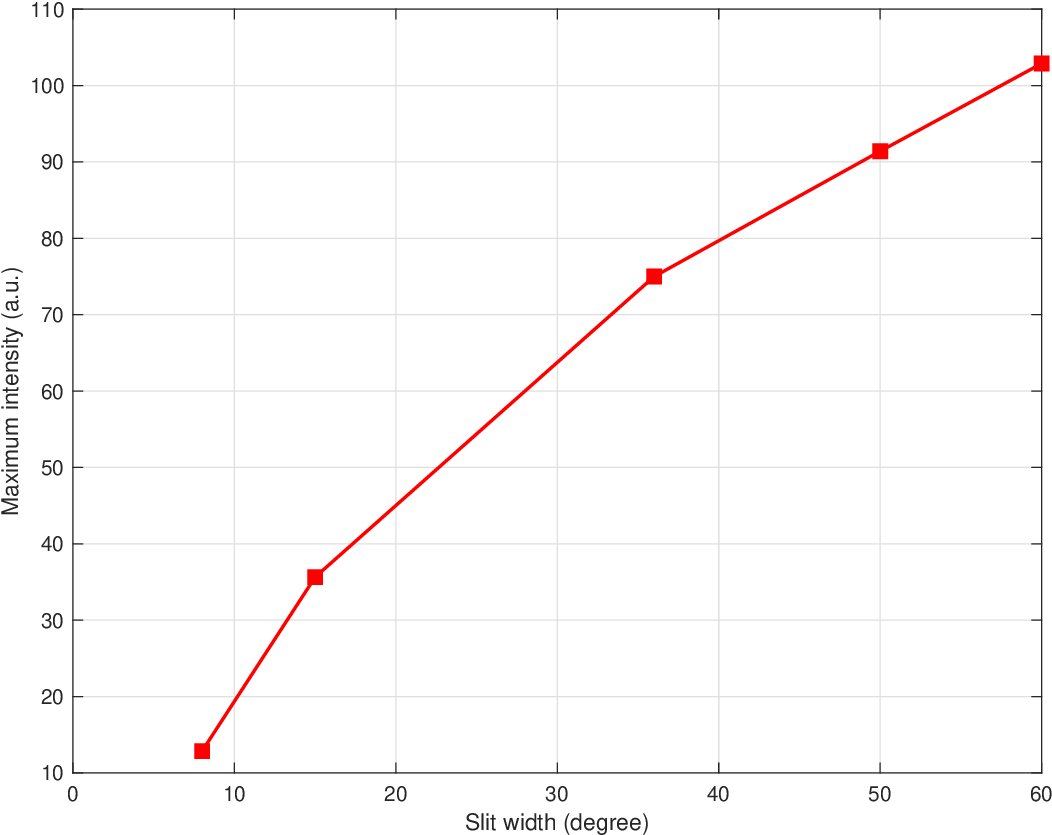}
\caption{Maximum intensity vs. $\alpha$}
\end{subfigure}
\hfill
\begin{subfigure}{0.48\linewidth}
\includegraphics[width=\linewidth]{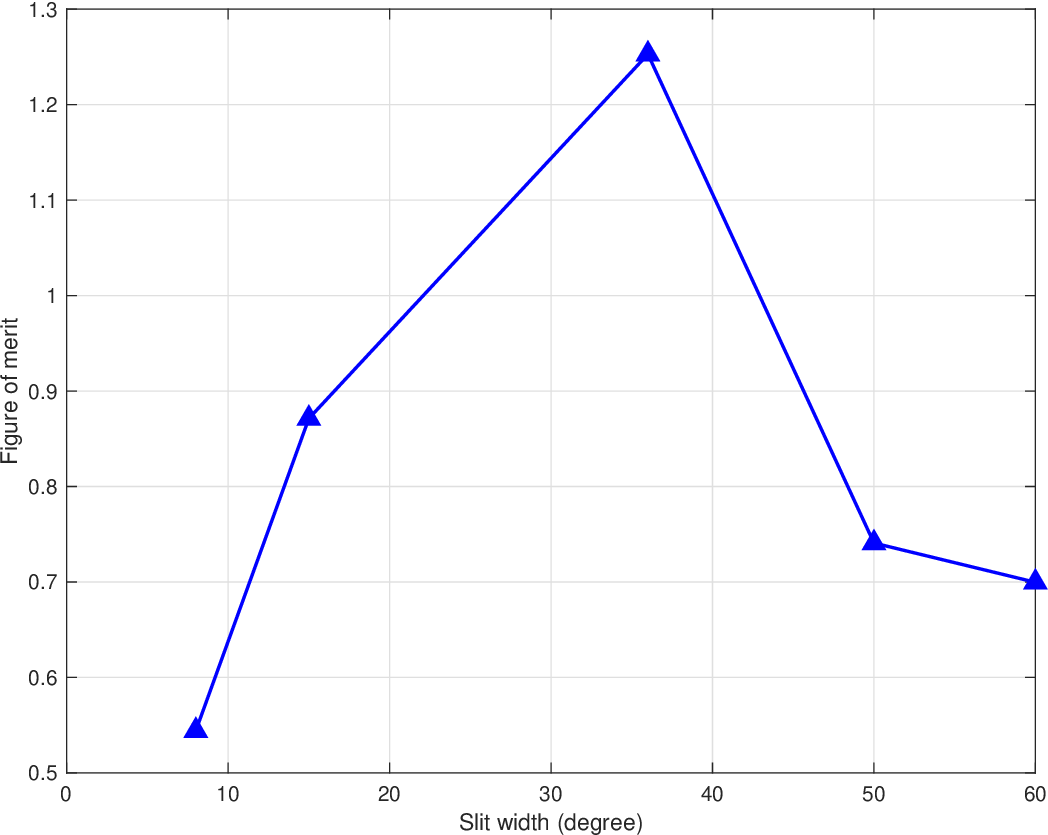}
\caption{FOM vs. $\alpha$}
\end{subfigure}
\caption{Quantitative analysis for $l=10$. Maximum intensity and FOM as functions of slit width. The FOM reaches its maximum at $\alpha=36^\circ$.}
\label{fig:quant_l10}
\end{figure}

\begin{figure}[htbp!]
\centering
\begin{subfigure}[b]{0.48\linewidth}
    \includegraphics[width=\linewidth]{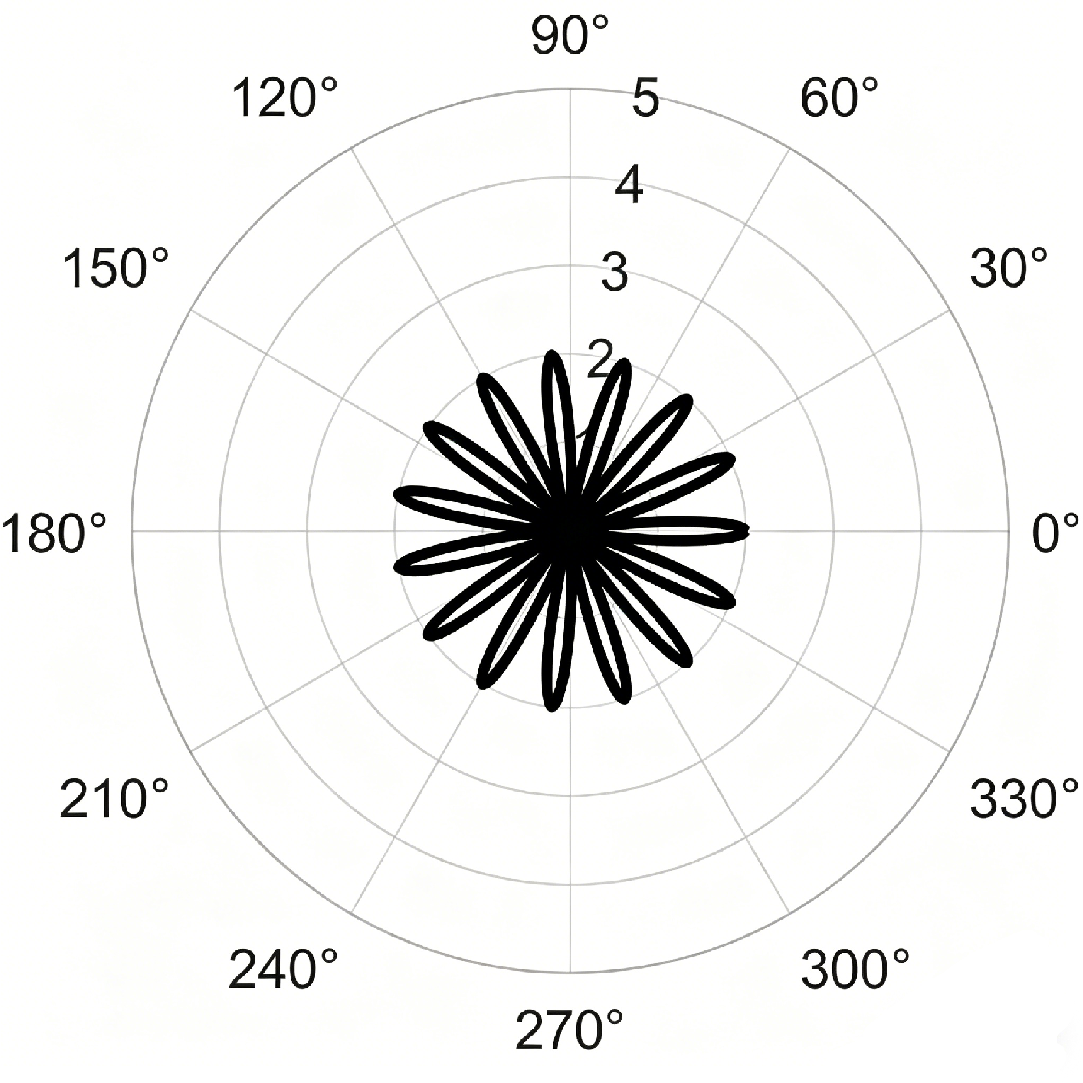}
    \caption{Theory, $\alpha=24^\circ$}
\end{subfigure}
\hfill
\begin{subfigure}[b]{0.48\linewidth}
    \includegraphics[width=\linewidth]{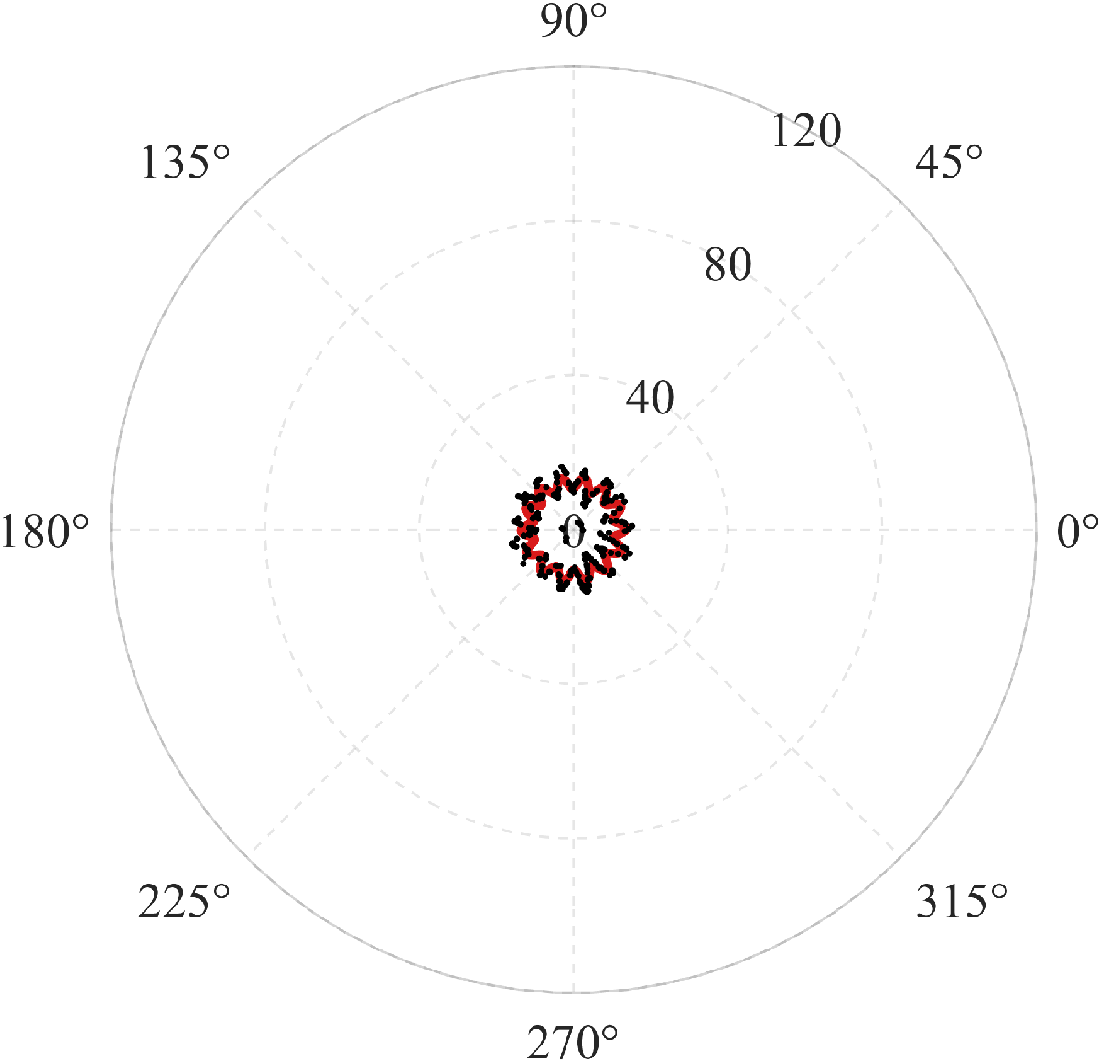}
    \caption{Exp., $\alpha=8^\circ$}
\end{subfigure}

\vspace{0.5em}
\begin{subfigure}[b]{0.48\linewidth}
    \includegraphics[width=\linewidth]{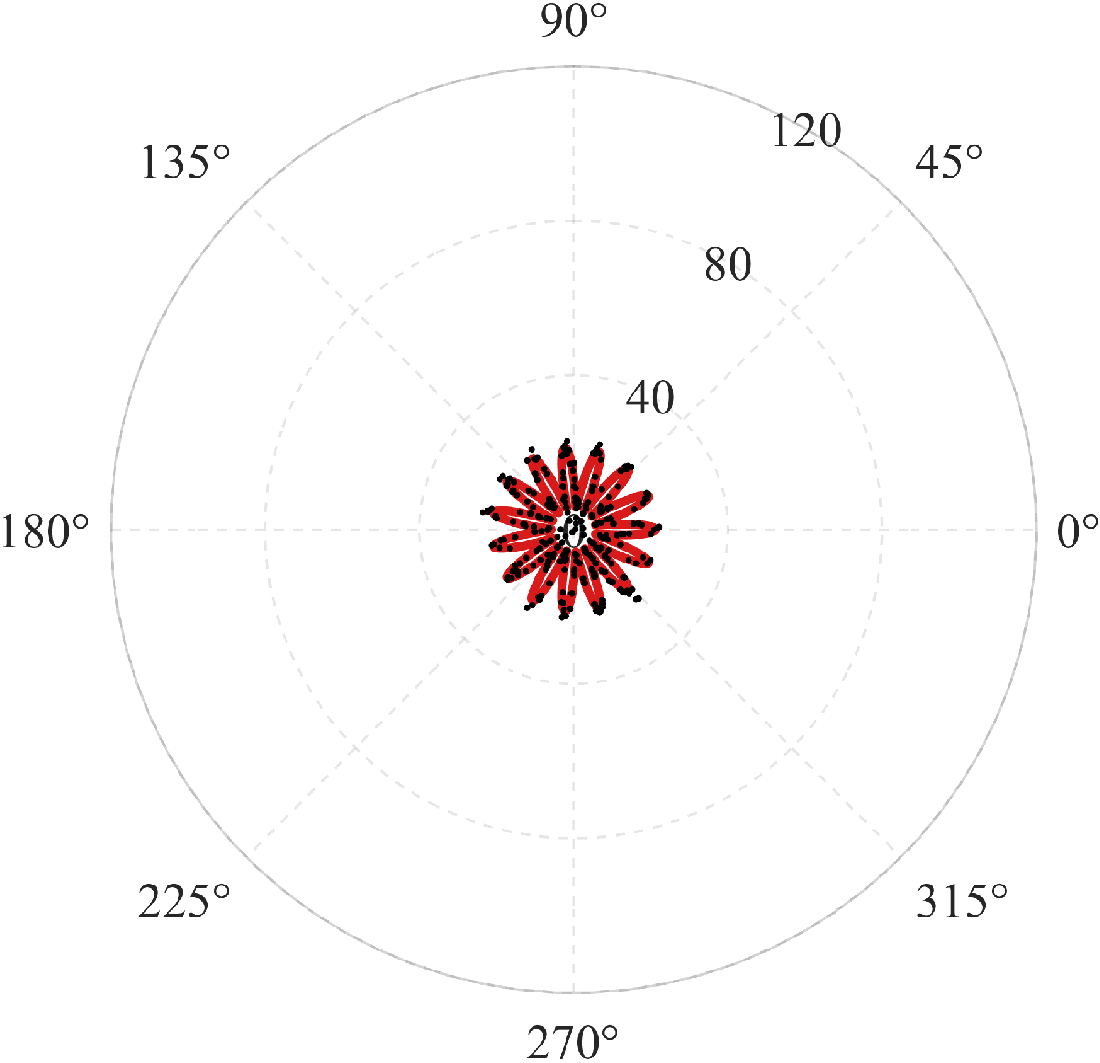}
    \caption{Exp., $\alpha=15^\circ$}
\end{subfigure}
\hfill
\begin{subfigure}[b]{0.48\linewidth}
    \includegraphics[width=\linewidth]{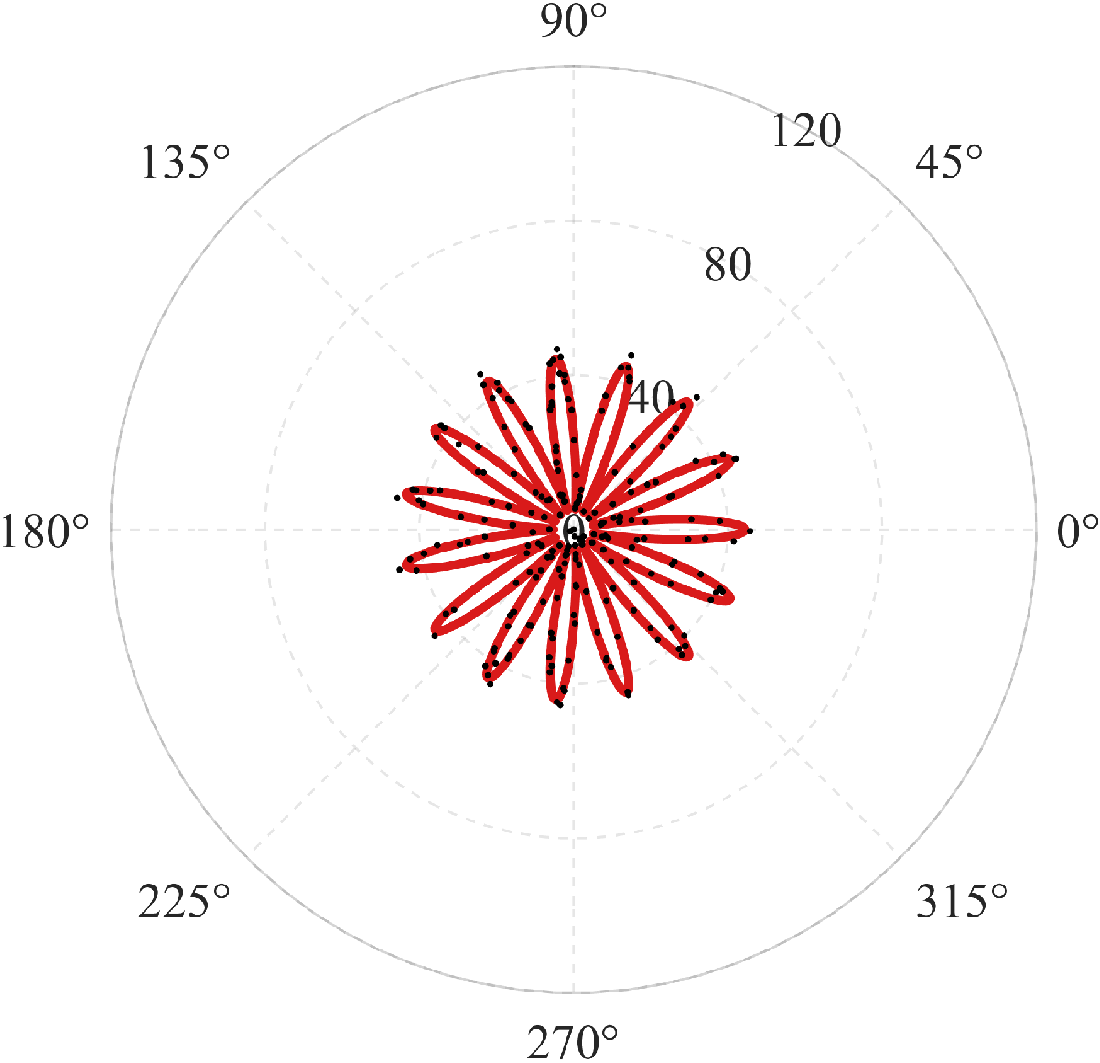}
    \caption{Exp., $\alpha=24^\circ$}
\end{subfigure}

\vspace{0.5em}
\begin{subfigure}[b]{0.48\linewidth}
    \includegraphics[width=\linewidth]{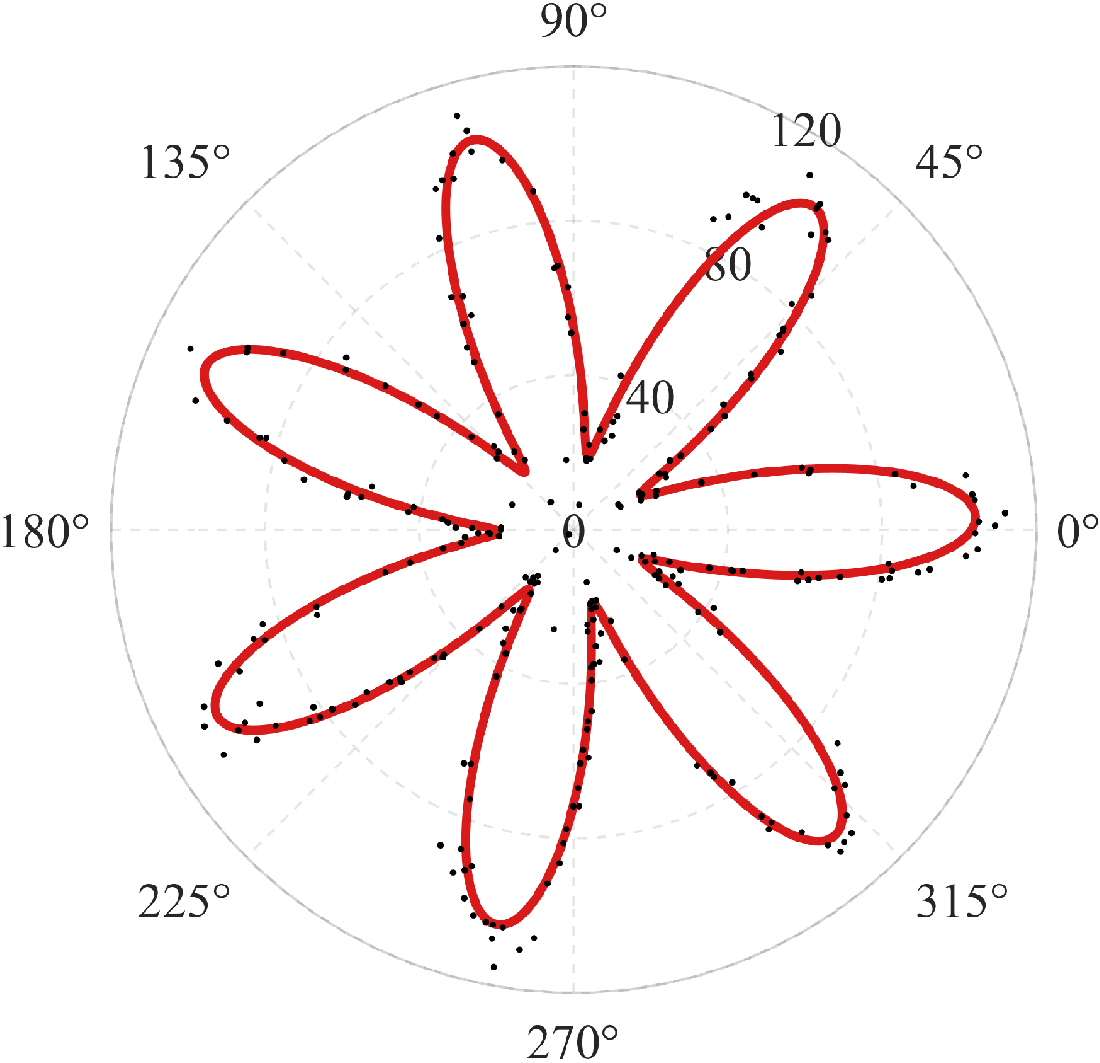}
    \caption{Exp., $\alpha=50^\circ$}
\end{subfigure}
\hfill
\begin{subfigure}[b]{0.48\linewidth}
    \includegraphics[width=\linewidth]{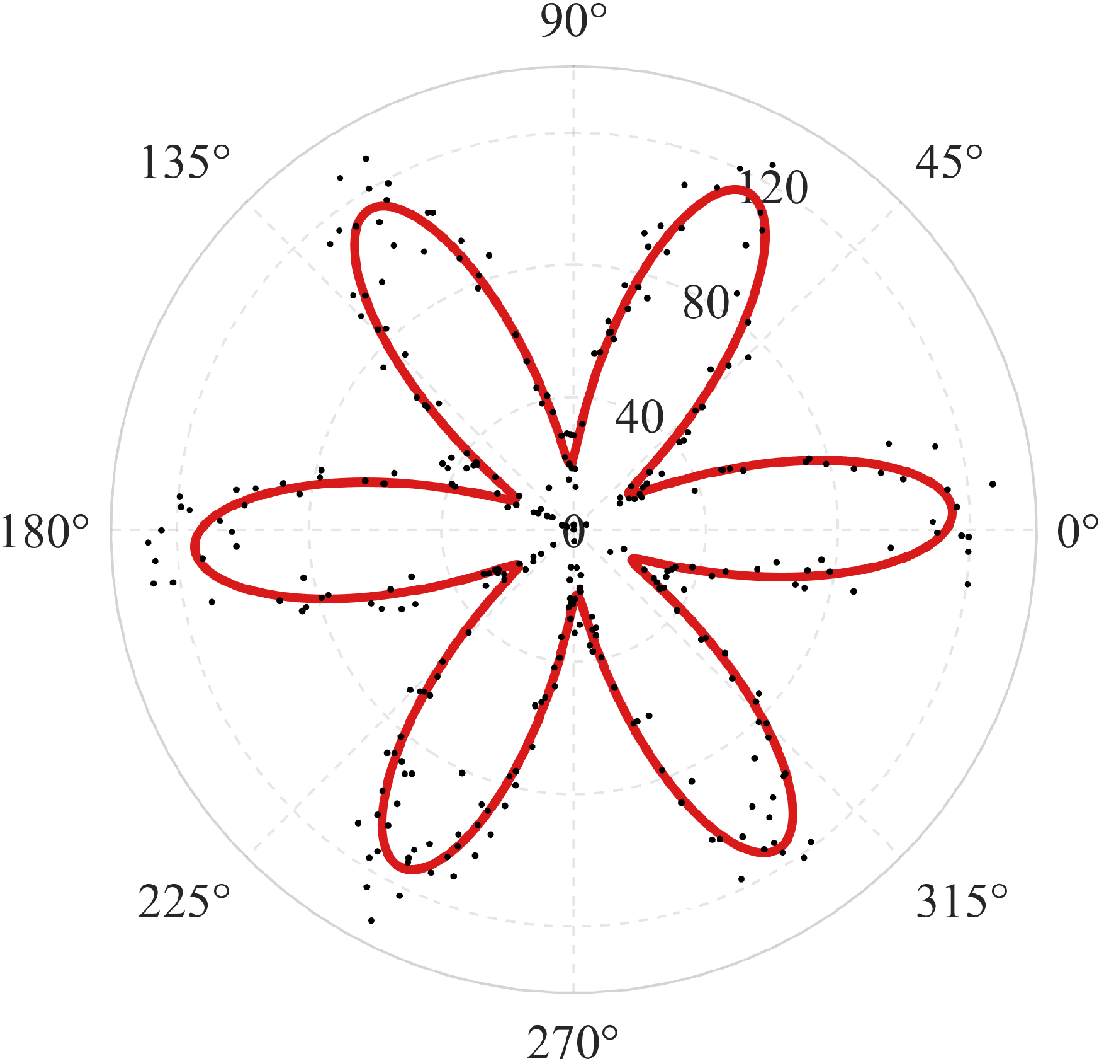}
    \caption{Exp., $\alpha=60^\circ$}
\end{subfigure}
\caption{Polar plots $I(\varphi)$ for $l=15$: (a) theoretical ideal curve at optimal slit width $24^\circ$; (b)--(f) experimental results. Optimal width produces fifteen clear lobes.}
\label{fig:polar_l15}
\end{figure}

\begin{figure}[htbp!]
\centering
\begin{subfigure}{0.48\linewidth}
\includegraphics[width=\linewidth]{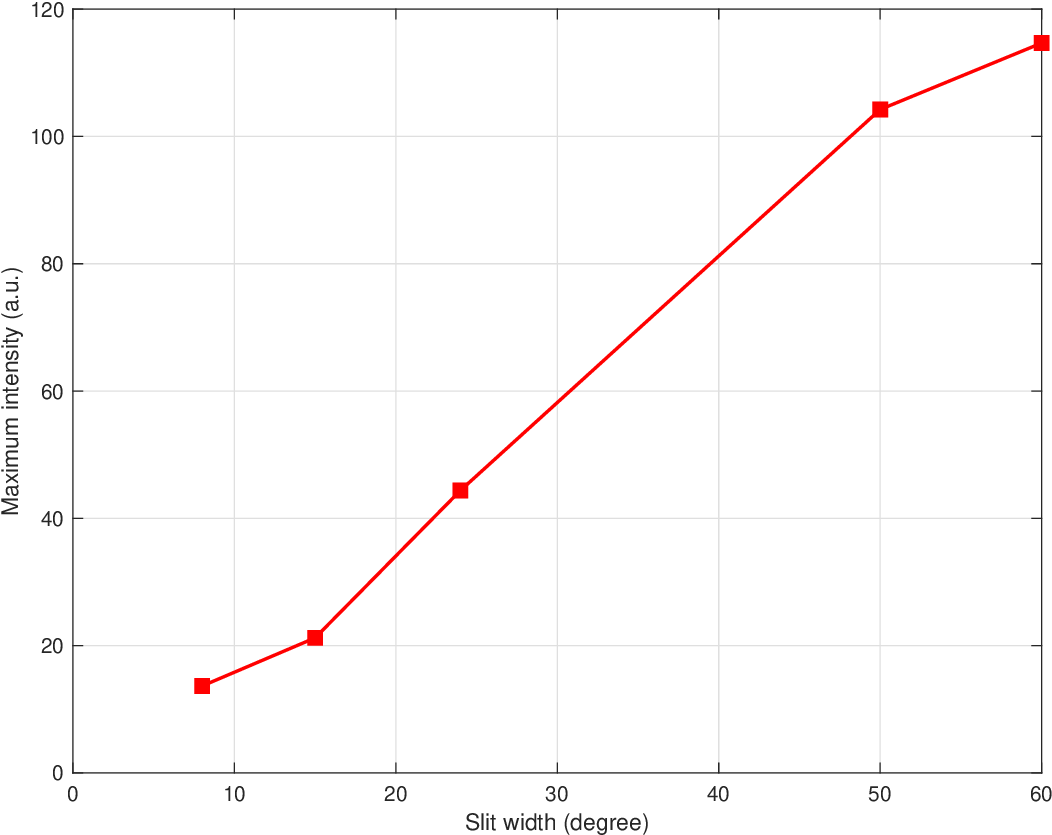}
\caption{Maximum intensity vs. $\alpha$}
\end{subfigure}
\hfill
\begin{subfigure}{0.48\linewidth}
\includegraphics[width=\linewidth]{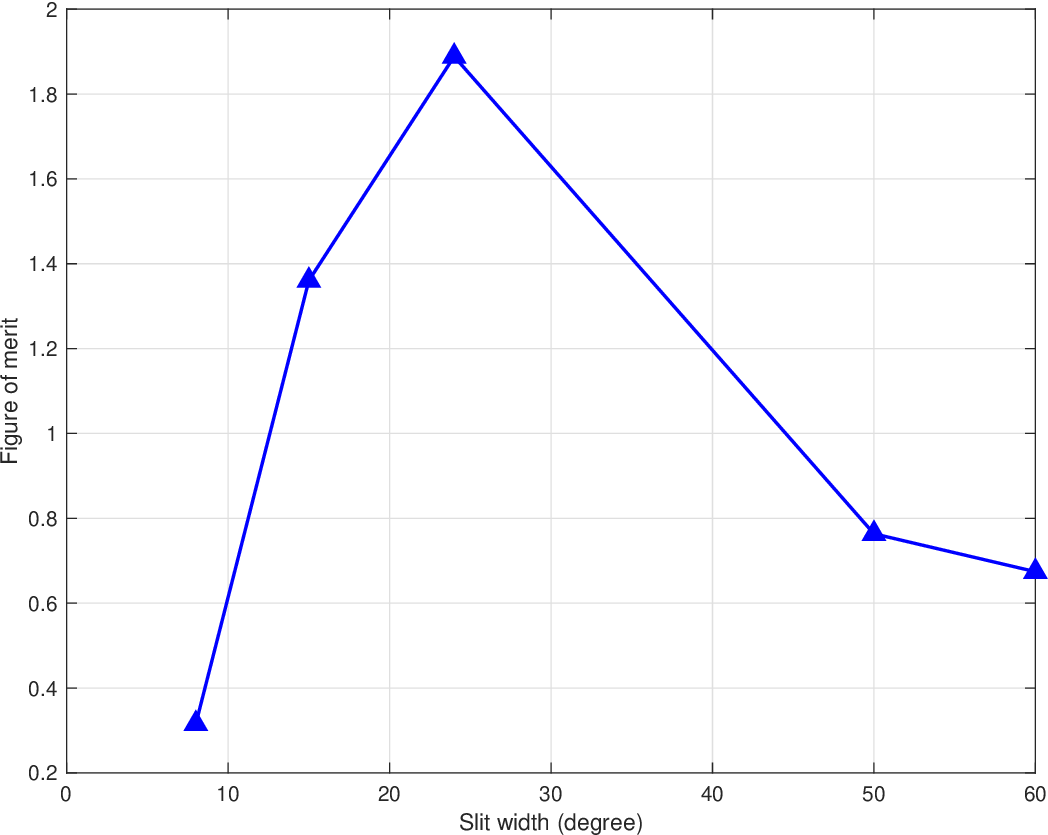}
\caption{FOM vs. $\alpha$}
\end{subfigure}
\caption{Quantitative analysis for $l=15$. The FOM is maximized at the optimal width $\alpha=24^\circ$.}
\label{fig:quant_l15}
\end{figure}

To identify the optimal slit width for topological charge measurement, multiple experiments were carried out. Here we present the results for $\ell = 5$, $10$, and $15$ as representative cases.Figures~\ref{fig:polar_l5}--\ref{fig:polar_l15} show the theoretical and experimental polar plots $I(\varphi)$ for different slit widths. The corresponding quantitative analyses (maximum intensity and figure of merit, FOM) are presented in Figs.~\ref{fig:quant_l5}--\ref{fig:quant_l15}. The FOM is defined as
\begin{equation}
\mathrm{FOM} = \frac{V}{\mathrm{FWHM}},
\label{eq:fom}
\end{equation}
where $V = (I_{\max}-I_{\min})/(I_{\max}+I_{\min})$ is the contrast and $\mathrm{FWHM}$ is the full width at half maximum of a single lobe. A larger FOM indicates that the interference signal maintains high contrast while exhibiting sharper peak shapes, which facilitates accurate determination of the topological charge.

\paragraph{Topological charge \(l = 5\).}
For \(l = 5\) (Fig.~\ref{fig:polar_l5}), the theoretical optimal slit width is \(72^\circ\). When \(\alpha = 8^\circ\) (much smaller than the phase period), the transmitted intensity was weak and the interference curve exhibited poor contrast, yielding a low FOM of 0.07. Increasing \(\alpha\) to \(50^\circ\) improved the signal, but the contrast remained inferior to the optimum. At the optimal width \(\alpha = 72^\circ\), the interference pattern displayed five clear lobes with maximum contrast; the FOM reached its peak value of 0.34. For \(\alpha = 80^\circ\) and \(90^\circ\) (exceeding the phase period), phase aliasing caused the loss of one lobe: only four peaks were observable, leading to an erroneous determination of \(|\ell|\). The corresponding FOM values dropped to 0.26 and 0.28, respectively. The evolution of the maximum intensity and FOM with slit width is shown in Fig.~\ref{fig:quant_l5}.

\paragraph{Topological charge \(l = 10\).}
For \(l = 10\) (Fig.~\ref{fig:polar_l10}), the optimal slit width is \(36^\circ\). At \(\alpha = 8^\circ\) and \(15^\circ\), the interference signal was too weak to reliably resolve the ten expected lobes. When \(\alpha\) was increased to the optimal \(36^\circ\), the polar plot exhibited ten well-defined lobes, the maximum intensity reached 74.99 (a.u.), and the FOM attained its highest value of 1.25. For \(\alpha = 50^\circ\), the number of lobes dropped to seven (FOM = 0.74); for \(\alpha = 60^\circ\), only six lobes remained (FOM = 0.70). Although the total intensity increased with larger slit widths, the loss of lobes would lead to a false topological charge reading. These results confirm that \(36^\circ\) is the optimal slit width for \(l = 10\).

\paragraph{Topological charge \(l = 15\).}
For \(l = 15\) (Fig.~\ref{fig:polar_l15}), the optimal slit width is \(24^\circ\). At suboptimal widths of \(8^\circ\) and \(15^\circ\), the FOM was only 0.31 and 1.36, respectively, and the lobes were poorly resolved. At the optimum \(\alpha = 24^\circ\), the polar plot displayed fifteen clear lobes with a maximum intensity of 44.38 and the highest FOM of 1.89. For \(\alpha = 50^\circ\) and \(60^\circ\), severe phase aliasing reduced the lobe count to seven and six, respectively, despite stronger overall intensity. The FOM decreased to 0.74 and 0.70, confirming that \(24^\circ\) is the optimal slit width for \(l = 15\).

\begin{figure}[htbp!]
\centering
\begin{subfigure}{0.48\linewidth}
\includegraphics[width=\linewidth]{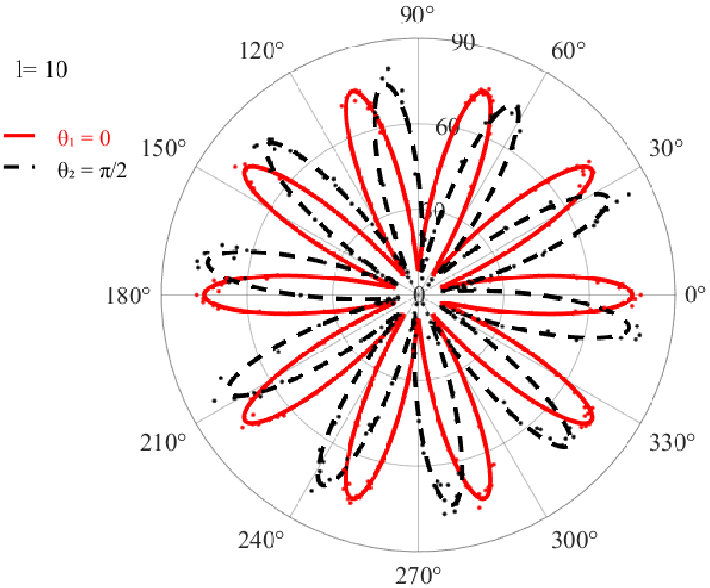}
\caption{$l=10$, $\alpha=36^\circ$}
\end{subfigure}
\hfill
\begin{subfigure}{0.48\linewidth}
\includegraphics[width=\linewidth]{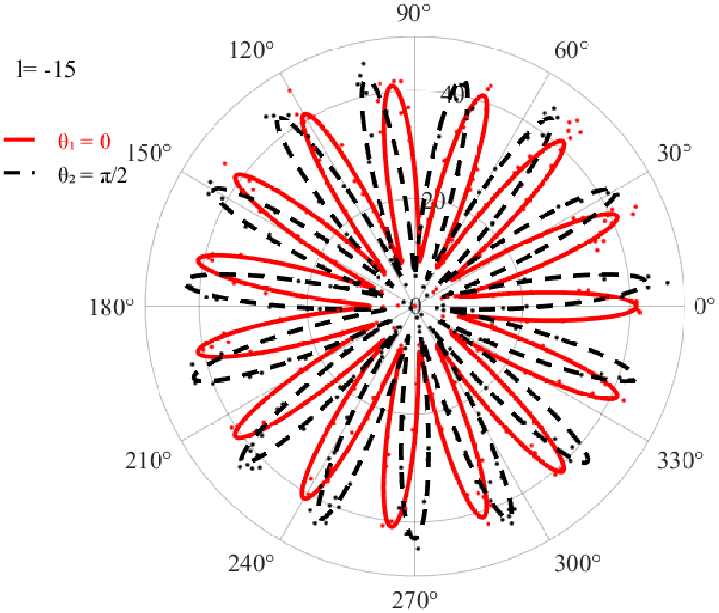}
\caption{$l=-15$, $\alpha=24^\circ$}
\end{subfigure}
\caption{$I(\varphi)$ curves with $\theta=0$ (black) and $\theta=\pi/2$ (red) for (a) $l=10$ and (b) $l=-15$ at optimal slit widths. Clockwise rotation in (a) confirms positive sign; counterclockwise rotation in (b) confirms negative sign.}
\label{fig:sign}
\end{figure}

\paragraph{Sign determination.}
The experiments described above measure only the magnitude of the topological charge. A complete characterization, however, requires determination of its sign. To this end, we introduced an additional phase shift of \(\theta = \pi/2\) on one slit.Figure~\ref{fig:sign} shows the measured \(I(\varphi)\) curves for \(l = 10\) and \(l = -15\) under the respective optimal slit widths. Here, we take \(l = 10\) and \(l = -15\) as examples to determine the sign of the topological charge. For \(l = 10\) (positive), the curve with \(\theta = \pi/2\) rotated clockwise relative to the \(\theta = 0\) curve. In contrast, for \(l = -15\) (negative), the \(\theta = \pi/2\) curve rotated counterclockwise. The rotation directions agree exactly with theoretical predictions, unambiguously revealing the sign of the topological charge.

Under the optimal slit width condition, the simulated and experimental curves agree excellently. Minor discrepancies, such as non-zero minima in the interference curves and slight angular offsets, are attributed to residual polarization impurities (extinction ratio 500:1), the finite fill factor (87\%) of the SLM. Importantly, these imperfections do not affect the number of lobes or the sign determination, demonstrating that the dynamic angular double-slit method is robust against typical experimental non-idealities.

\subsection{Optimal slit width condition}

\begin{figure}[htbp]
\centering
\begin{subfigure}[b]{0.3\linewidth}
\includegraphics[width=\linewidth]{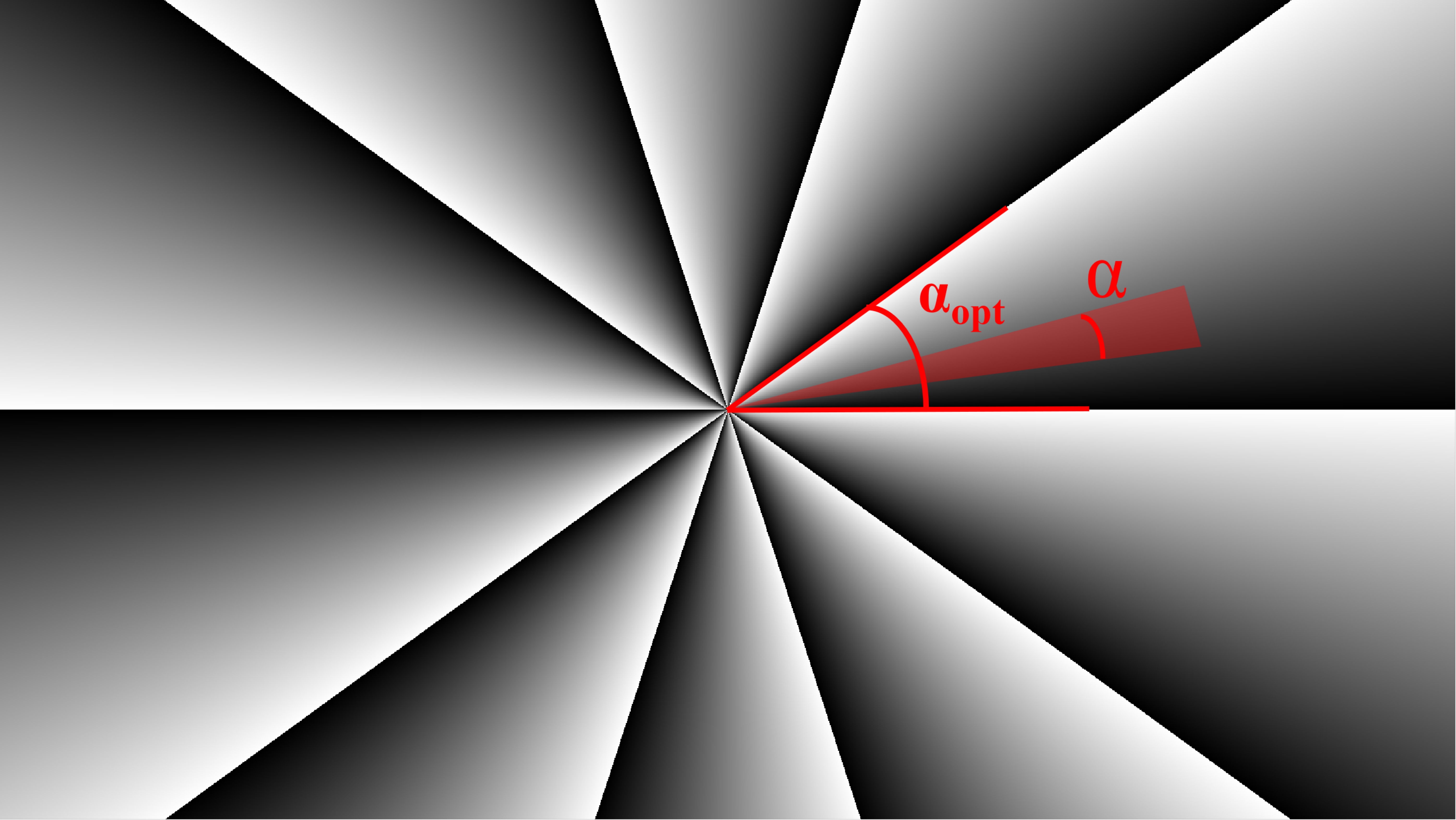}
\caption{$\alpha < 2\pi/|l|$}
\label{fig:slit_small}
\end{subfigure}
\hfill
\begin{subfigure}[b]{0.3\linewidth}
\includegraphics[width=\linewidth]{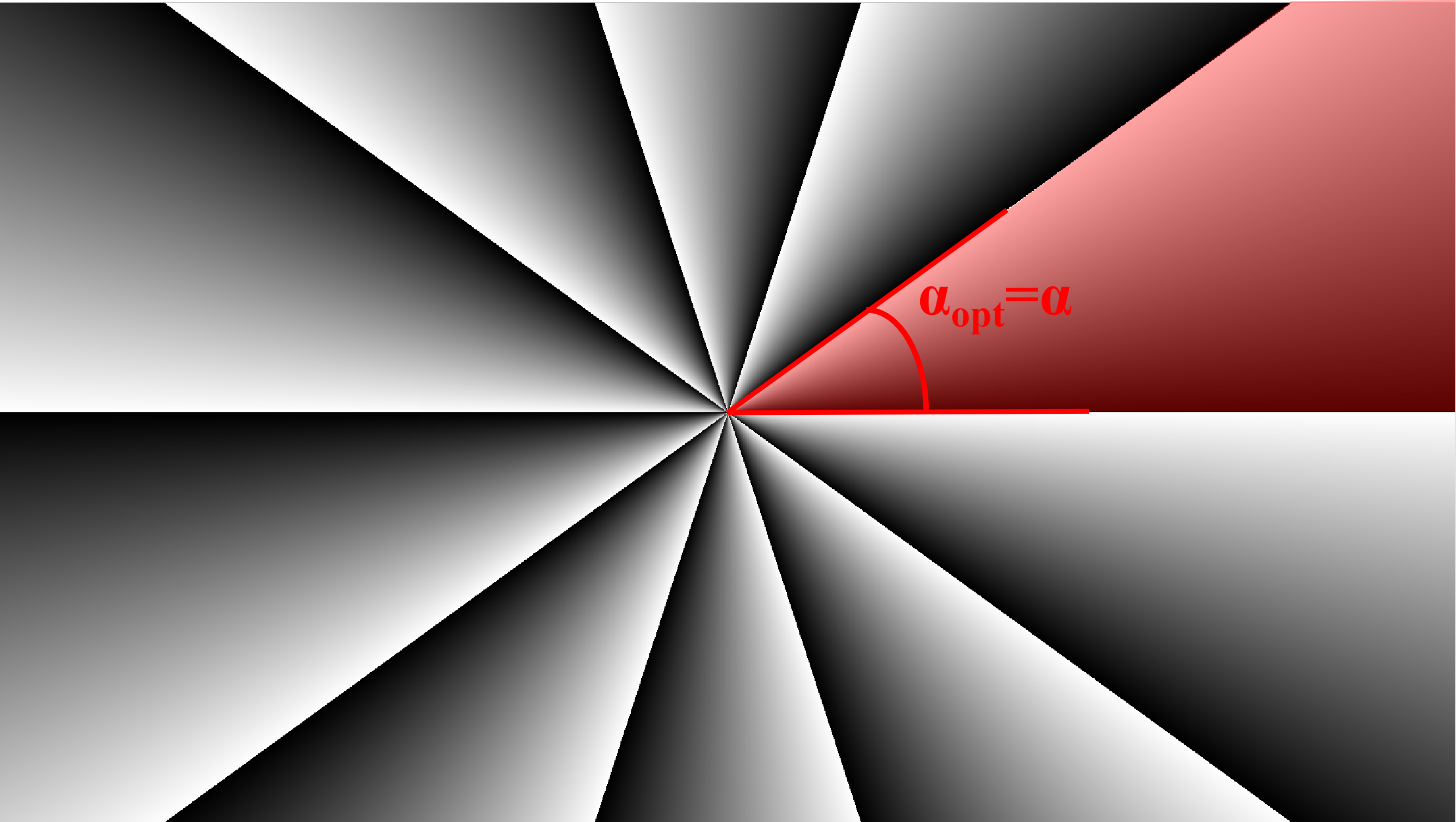}
\caption{$\alpha = 2\pi/|l|$}
\label{fig:slit_equal}
\end{subfigure}
\hfill
\begin{subfigure}[b]{0.3\linewidth}
\includegraphics[width=\linewidth]{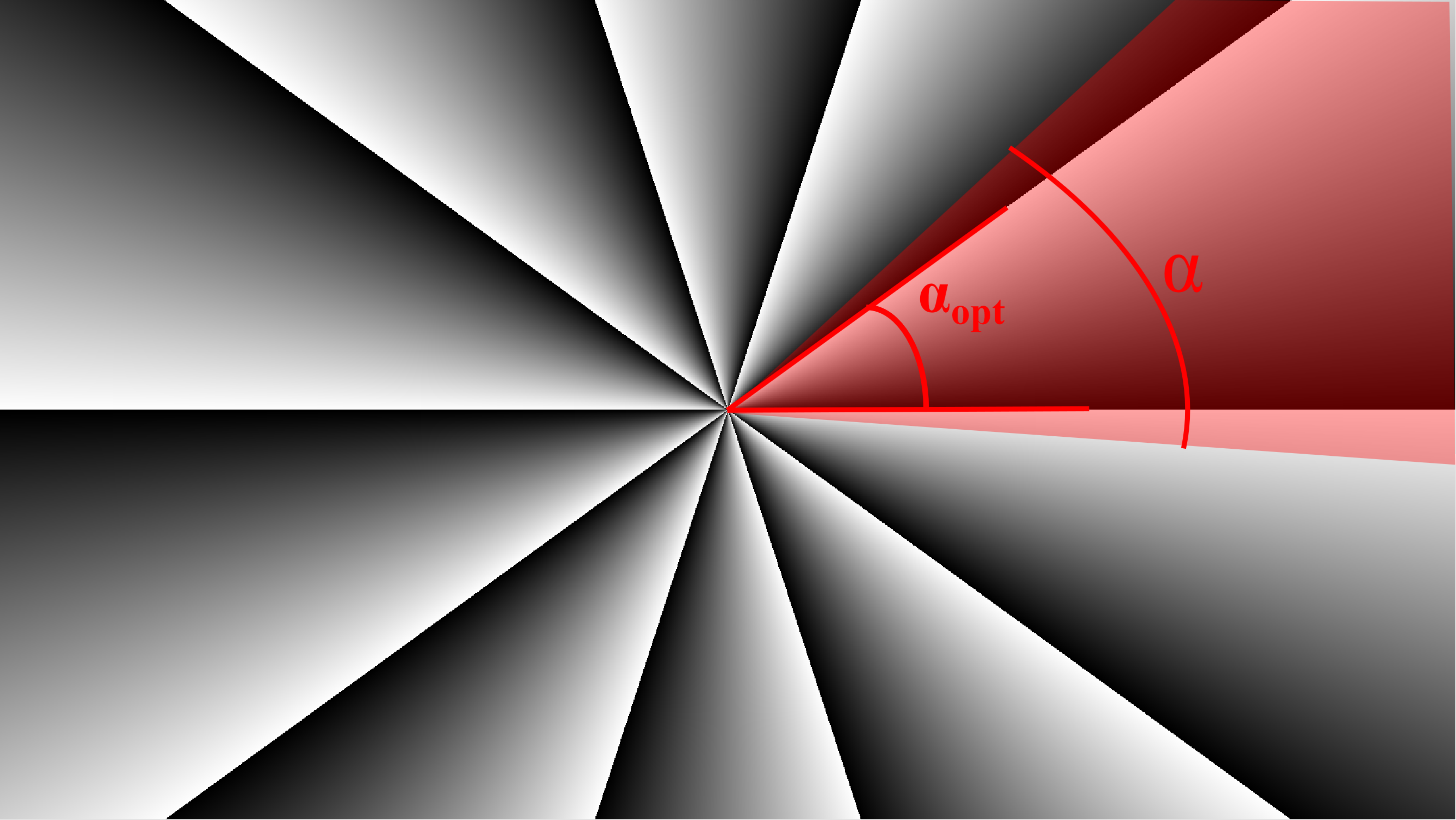}
\caption{$\alpha > 2\pi/|l|$}
\label{fig:slit_large}
\end{subfigure}
\caption{Illustration of phase sampling for a vortex beam with $l=10$ under three slit width conditions. (a) Slit width smaller than the phase period, causing phase truncation. (b) Slit width exactly equal to the phase period, capturing one full $2\pi$ cycle (optimal). (c) Slit width larger than the phase period, leading to phase aliasing.}
\label{fig:phase_conditions}
\end{figure}

Based on the experimental and simulated results presented above for $\ell = 5, 10,$ and $15$, we can summarize the critical condition for achieving the highest measurement accuracy. A vortex beam possesses a helical phase front that varies linearly with the azimuthal angle $\phi$ at a rate determined by the topological charge $\ell$, with a phase period of $2\pi/|\ell|$. The angular width $\alpha$ of each sector-shaped slit determines the range of phase covered by that slit. The selection of $\alpha$ critically affects the completeness of phase sampling and thus the measurement accuracy. Depending on the relationship between $\alpha$ and the phase period, three typical cases can be distinguished, as illustrated in Figs.~\ref{fig:phase_conditions}(a), (b), and (c) for a vortex beam with $\ell = 10$.

\textbf{Case 1: $\alpha < 2\pi/|\ell|$ (phase truncation).} ... (rest of the text)

\section{Conclusion}

We have systematically investigated the critical role of slit width in dynamic angular double-slit interferometry for measuring the topological charge (TC) of vortex beams. Both theoretically and experimentally, we established the optimal condition $\alpha_{\mathrm{opt}} = 2\pi/|\ell|$, under which each slit samples exactly one complete $2\pi$ phase period. This condition maximizes the fidelity of the interference signal, yielding the highest figure of merit. Experiments for $\ell = 5$, $10$, and $15$ confirmed the universality of this criterion. Furthermore, by introducing a $\pi/2$ phase shift, we demonstrated unambiguous sign determination for both $\ell = 10$ and $\ell = -15$.

It should be noted that the optimal condition is applicable for $|\ell| \ge 2$. For $|\ell| = 1$, the required slit width would be $360^\circ$, which degenerates the angular double-slit geometry into two full rings, making the method unsuitable. For unknown TCs, an iterative coarse-to-fine measurement can be adaptively implemented using an SLM.

Our findings provide a clear and practical guideline for high-precision OAM characterization. This simple, robust method holds great potential for applications in optical communication, micromanipulation, and quantum information processing where accurate OAM metrology is paramount \cite{Yao2011, Shen2019}.

\backmatter

\bmsection{Funding} Natural Science Foundation of Chongqing, China (CSTB2022NSCQ-MSX0428); Through Train Project for Ph.D. of Chongqing (CSTB2022BSXM-JSX0004); Fundamental Research Funds for the Central Universities (2021CDJQY-041).

\bmsection{Disclosures} The authors declare no conflicts of interest.

\bmsection{Data availability} Data underlying the results presented in this paper are available from the authors upon reasonable request.

\end{document}